\documentclass[twocolumn,twocolappendix]{aastex631}

\usepackage{amsmath} 
\usepackage{ulem}

\begin{document}
\title{Formation and Early Evolution of Protoplanetary Disks under Nonuniform Cosmic-Ray Ionization}

\author[0009-0007-8872-4044]{Erika Nishio}
\affiliation{Astronomical Institute, Graduate School of Science, Tohoku University, Sendai, Miyagi 980-8578, Japan}

\author[0000-0001-8105-8113]{Kengo Tomida}
\affiliation{Astronomical Institute, Graduate School of Science, Tohoku University, Sendai, Miyagi 980-8578, Japan}

\author[0000-0003-0548-1766 ]{Yuki Kudoh}
\affiliation{Astronomical Institute, Graduate School of Science, Tohoku University, Sendai, Miyagi 980-8578, Japan}

\author[0000-0003-2579-7266]{Shigeo S. Kimura}
\affiliation{Astronomical Institute, Graduate School of Science, Tohoku University, Sendai, Miyagi 980-8578, Japan}
\affiliation{Frontier Research Institute for Interdisciplinary Sciences, Tohoku University, Sendai, Miyagi 980-8578, Japan}



\begin{abstract}
Angular momentum transport by magnetic fields is important for formation and evolution of protoplanetary disks. The effects of magnetic fields are suppressed due to non-ideal magnetohydrodynamic (MHD) effects such as ambipolar diffusion and Ohmic dissipation, which depend on the degree of ionization. Cosmic rays (CRs) are the primary source of ionization in star-forming clouds, and their distribution is nonuniform as it is affected by gas density and magnetic fields. Therefore, CRs, magnetic fields, and gas interact with each other. In this work, we develop a new fully implicit cosmic ray transport module in Athena++ and perform three-dimensional simulations of disk formation from collapse of molecular cloud cores. Since CRs are strongly attenuated in the dense gas at the disk scale, distribution of magnetic fields is considerably altered compared to conventional models assuming a uniform ionization rate. While the total magnetic fluxes accreted onto the disks remain similar as the gas outside the disks remain sufficiently ionized and well coupled, the magnetic fields in the disks are less twisted due to the stronger non-ideal MHD effects. As a consequence, magnetic angular momentum transport is strongly suppressed at the disk scale, resulting in more gravitationally unstable disks with more prominent spiral arms. Our simulations demonstrate influence of non-uniform ionization resulting from CR transport and attenuation on the disk formation and evolution.
\end{abstract}



\section{Introduction} \label{intro}
Since protoplanetary disks provide initial and boundary conditions for planet formation and supply mass and angular momentum to protostars, they are key objects for understanding star and planet formation processes. Magnetic fields play an important role in formation and evolution of protoplanetary disks (\cite{Tsukamoto2023_PP7}: Chapter 9 in Protostars and Planets VII reviews). For example, magnetic braking efficiently removes angular momentum from a protoplanetary disk, making its size smaller \citep{Basu1994,Tomisaka2000,Hennebelle2008,Mellon2009,Machida2011}. In addition, interaction between magnetic fields and rotating gas drives mass ejection phenomena known as outflows and jets \citep{Blandford1982,Tomisaka1998,Machida2013,Machida2014}. These angular momentum and mass transport mechanisms due to magnetic fields are affected by nonideal MHD effects such as Ohmic dissipation, Hall effect, and ambipolar diffusion \citep{Wardle2007, Machida2007,Tomida2015,Tsukamoto2015,Tsukamoto2015_Hall,Masson2016,Wurster2021}. While Ohmic dissipation and ambipolar diffusion are diffusive processes that suppress angular momentum transport by magnetic fields, the Hall effect can enhance or suppress magnetic angular momentum transport depending on the relative directions between the angular momentum and magnetic fields.

The strength of the nonideal MHD effects depends on the ionization degree of gas; the lower the ionization degree is, the stronger the effects are. The energy of $15.6 \,\mathrm{eV}$ is required to ionize $\mathrm{H_2}$, but ultraviolet (UV) photons that ionize $\mathrm{H_2}$ are easily shielded in star-forming regions with column density as large as $\Sigma \gtrsim 10^{-3}\,\mathrm{g\,cm^{-2}}$ \citep{Bergin2007} and cannot penetrate into molecular cloud cores. Therefore, external UV photons are not very effective in ionization unless there is a massive star nearby. The main sources of ionization in the early phase of star formation in molecular cloud cores are low-energy cosmic rays (CRs) and decay of radioactive nuclides.

The ionization rate per $\mathrm{H_2}$ particle $\zeta \,\mathrm{(s^{-1})}$ in molecular clouds and protoplanetary disks is highly uncertain and/or diverse in different environments. Spitzer's ionization rate, $\zeta = 10^{-17}\mathrm{\,s^{-1}}$ \citep{Spitzer1978}, is often used as a fiducial value in interstellar media (ISM), but recent observations using molecular lines such as $\mathrm{N_2H^+,\, HCO^+,\,H^{13}CO^+,\,}$ and $\mathrm{DCO^+}$ indicate that the ionization rates in molecular clouds largely vary in the range of $10^{-15}-10^{-18}\mathrm {\,s^{-1}}$. For example, \citet{Pineda2024} reported excess of the ionization rate around protostars in an active star-forming region NGC1333, which is as high as $10^{-15}-10^{-16}\,\mathrm{s^{-1}}$. In the disk scales, observations are somewhat controversial. The ionization rate around a Class-0 protostar B335 is as high as $10^{-13}-10^{-15}\mathrm{\,s^{-1}}$ \citep{Cabedo2023}. On the other hand, the ionization rate in the disk midplane around TW Hya is as low as $\zeta < 10^{-19}\,\mathrm{s^{-1}}$ \citep{Cleeves2015}. The low ionization rate is also supported from the narrow ring-gap structures which infer weak turbulence in the disk \citep{Pinte2016}. In addition, the ionization rate is not spatially constant even within a single disk \citep{Seifert2021,Long2024}. These observations suggest that the ionization rates are not spatially uniform nor temporally constant during the evolution in star-forming regions.

There are three possible ionization sources in star-forming clouds \citep{Padovani2013b,Padovani2014,Padovani2018,Silsbee2018,Silsbee2019,Takasao2019,Margot2021,Kimura2023}: (1) Galactic CRs produced by supernovae and propagate into star-forming regions, (2) low-energy CRs and gamma-rays emitted from decay of radioactive nuclides, and (3) internally-produced CRs which are accelerated at shocks and/or magnetic reconnection events around central protostars.

Galactic CRs penetrate from outside molecular cloud cores into the interior along magnetic field lines, losing the energy of CRs due to collisions with gas. This attenuation of the CR energy is significant where the column density exceeds $\Sigma \gtrsim 96 \mathrm{\,g\,cm^{-2}}$ \citep{Umebayashi2009}. When the magnetic field lines are twisted, the effective column density gets larger and CRs are more attenuated \citep{Fujii2022}. In star-forming molecular cloud cores, magnetic field lines tend to converge toward the center as a result of gravitational collapse.  In such a configuration, the magnetic focusing and mirroring effects affect the CR distribution as well. These effects are more or less equally effective, and the CR energy density is reduced by about a factor of 1/2 from the interstellar value besides the attenuation \citep{Silsbee2019,Fujii2022}.

In regions of high column density such as the midplane of a protoplanetary disk, low-energy CRs and gamma-rays emitted from radioactive radionuclides contribute to ionization. Because of the low energies, these particles do not propagate very far, and this can be considered as a local process. Short-lived radioactive nuclides (SLR; the most relevant one is $^{26}\mathrm{Al}$ whose half-life is $\mathrm{7.4\times 10^5\;yr}$) contribute more to ionization than long-lived radioactive nuclides (LLR; e.g., $\mathrm{^{40}K}$ with half-life of $\mathrm{1.3\times 10^9\;yr}$), but decay faster. While LLR is more or less ubiquitous in the Galactic environments, the SLR abundance should vary from place to place. Assuming the abundance of $\mathrm{^{26}Al}$ in the early solar system estimated from meteorite analyses, SLR can maintain the ionization rate of $\mathrm{(7-10)\times 10^{-19}\;s^{-1}}$ for $\mathrm{1\,Myr}$ \citep{MacPherson1995,Umebayashi2009}.

After a central protostar is formed, it emits ionizing radiation such as UV and X-rays. In addition, shocks in the accreting flow and magnetic reconnection events around the protostar can accelerate CRs and contribute to ionization. \cite{Padovani2016} investigated Diffusive Shock Acceleration (DSA:\cite{Bell1978,Drury1983}) in shock waves around a protostar. They found that protons can be accelerated to relativistic energies by shock waves in jets and by accretion shocks on a stellar surface. Another possible CR acceleration sites are protostellar flares, where protons can be accelerated up to TeV energies at shocks formed by interaction between the stellar magnetic field and reconnection outflows \citep{Takasao2019,Kimura2023}.

Previous numerical studies showed that the size of protoplanetary disks depends on the ionization rate\citep{Li2011,Kuffmeier2020,Kobayashi2023}. \cite{Kuffmeier2020} found that when the ionization rate is as high as $10^{-16}\mathrm{\,s^{-1}}$, angular momentum transport by magnetic braking works efficiently and a disk shrinks and eventually disappears, while the magnetic braking is well suppressed and a large disk is formed in the case of a low ionization rate of $10^{-17}\mathrm{\,s^{-1}}$. These results indicate that the ionization rate has a significant impact on the disk evolution and structure.

Previous studies often assume the ionization rate to be constant both in space and time or model it as a simple function of the column density \citep{Kunz2009, Kunz2010}. However, because CRs, gas and magnetic fields interact mutually, it is necessary to model these processes consistently in order to understand the effect of non-uniform ionization rate in space and time on formation and evolution of protoplanetary disks. In this work, we present results of non-ideal MHD simulations using a newly developed CR transport solver. As the first step in our study of star formation including the CR transport, we study the early stage of star formation from collapse of molecular cloud cores to the formation of protostellar cores. We consider ionization by the Galactic CRs and decay of radioactive nuclides, and leave the internally-produced CRs for future works because they become important in the later stages. This paper is organized as follows. We explain the equations and initial conditions in Section \ref{Method}, and present the results of the numerical simulations in Section \ref{Result}. We discuss in Section \ref{Discussion} and summarize conclusions in Section \ref{Conclusion}. 

\section{Methods and Models} \label{Method}
\subsection{Basic Equations and Methods}
To study formation of protoplanetary disks considering the mutual evolution of CRs and MHD, we perform 3D simulations using a public non-ideal MHD simulation code \texttt{Athena++} \citep{Stone2020} with a newly-developed CR transport module. The governing equations including non-ideal MHD and self-gravity are given as
\begin{align}
&\frac{\partial \rho}{\partial t}+\nabla \cdot (\rho \boldsymbol{v}) =0, \\
\begin{split}
&\frac{\partial \rho\boldsymbol{v}}{\partial t} + \nabla \cdot\left[ \rho\boldsymbol{v\otimes v} 
+\left(p+\frac{|\boldsymbol{B}|^2}{8\pi}\right)\mathbb{I}-\frac{1}{4\pi}\boldsymbol{B \otimes B}\right]\\
&=-\rho \nabla \Phi ,
\end{split}\\
&\nabla^{2} \Phi=4 \pi G \rho,\\
&\frac{\partial \boldsymbol{B}}{\partial t}-\nabla \times\left[(\boldsymbol{v} \times \boldsymbol{B})-\eta_\Omega \boldsymbol{J}-\frac{\eta_{\mathrm{AD}}}{|\boldsymbol{B}|^2}\boldsymbol{B} \times(\boldsymbol{J} \times \boldsymbol{B})\right]=0,\label{nonidealBeq}\\
&\boldsymbol{J} \equiv \nabla \times \boldsymbol{B},\\
&\nabla \cdot \boldsymbol{B}=0,
\end{align}
where $\rho ,\,\boldsymbol{v},\,p,\,\boldsymbol{B}, \eta_\Omega, \eta_{\mathrm{AD}}$ and $\Phi$ are the gas density, gas velocity, gas pressure, magnetic flux density, Ohmic resistivity, ambipolar diffusion coefficient, and gravitational potential, respectively, and  $G=6.673 \times 10^{-8} \mathrm{~cm}^3 \mathrm{~g}^{-1} \mathrm{~s}^{-2}$ is the gravitational constant. Note that both the Ohmic resistivity, $\eta_\Omega$, and the ambipolar diffusion coefficient, $\eta_\mathrm{AD}$, are expressed in units of $\mathrm{cm^2\,s^{-1}}$.

Instead of solving the energy equation, the temperature is computed using the barotropic approximation \citep{Zhao2018M} modeling the thermal evolution of the central gas element in a collapsing cloud obtained from a radiation hydrodynamic simulation \citep{Tomida2013}. 
{\small
\begin{align}
T= 
\begin{cases}
T_0+1.5 \left(\frac{\rho}{10^{-13}}\right) \, & (\rho<10^{-12})\\ 
\left(T_0+15\right)\left(\frac{\rho}{10^{-12}}\right)^{0.6}\, & (10^{-12} \leq \rho < 10^{-11} )\\
10^{0.6}\left(T_0+15\right)\left(\frac{\rho}{10^{-11}}\right)^{0.44} \, & (10^{-11} \leq \rho)
\end{cases},
\end{align}}
where the units of the density and temperature are $\mathrm{g\,cm^{-3}}$ and $\mathrm{K}$, and $T_0 = 10$. 

The Ohmic resistivity $\eta_{\Omega}$ and ambipolar diffusion coefficient $\eta_{\mathrm{AD}}$ are given from pre-calculated tables as functions of $\rho,\,T,\,|\boldsymbol{B}|,$ and ionization rate $\zeta$ \citep{Okuzumi2009}. We assume the dust size of $0.1\mathrm{\;\mu m}$ and the dust-to-gas ratio of 0.01. The ionization rate $\zeta$ is calculated from the CR energy density given by the CR transport equation (see below). In addition, we include thermal ionization of Potassium which provides sufficient electrons to make the gas coupled to magnetic fields when the gas temperature exceeds $T\sim 1,000\,\mathrm{K}$. Note that the Hall effect is not included for simplicity.

We calculate the ideal MHD part using the second-order piecewise linear reconstruction and second-order van-Leer time integrator with the HLLD approximate Riemann solver \citep{Miyoshi2005} and Constrained Transport method \citep{Gardiner2008}. The non-ideal MHD part is solved using the operator-split technique, and accelerated with the super-time-stepping method \citep{Alexiades1996,O'Sullivan2006,O'Sullivan2007,Meyer2014}. We use the full multigrid solver \citep{Tomida2023} for solving the Poisson equation of the self-gravity.

For the CR transport, moment-based equations are often used instead of the underlying kinetic equation in order to save computational cost. In this work, we adopt the diffusion approximation which solves only the zeroth moment equation. While the diffusion approximation may not be sufficiently accurate for the CR transport in star-forming clouds, we adopt it in order to circumvent some computational difficulties. First, the time scales for the CR transport and interaction with gas can be very short. Therefore, we need to adopt some numerical techniques to enable efficient computation. One option is to solve the zeroth and first moment equations using an explicit time integrator combined with the reduced-speed-of-light approximation. Although such a solver is available in \texttt{Athena++} \citep{Jiang2018}, we have decided not to use it because the reduced-speed-of-light approximation is not accurate for the problem of our interest, which has an enormous dynamic range (more than ten orders of magnitude in the gas density) and contains both highly ``cosmic-ray" thick and thin regions \citep{Skinner2013,Hopkins2022}. Therefore, we adopt the diffusion approximation and solve it using a fully implicit time integrator instead. We discuss the limitation of the scheme and possible impact on the results later in Section~\ref{Sec:DiffusionApproximation}.

The CR transport equation with the diffusion approximation reads
\begin{align}
    &\frac{\partial}{\partial t} e_{\mathrm{c}}+\nabla \cdot\left(e_{\mathrm{c}} \boldsymbol{v}\right)-p_{\mathrm{c}} \nabla \cdot \boldsymbol{v}=\nabla \cdot\left(\mathbb{K} \nabla e_{\mathrm{c}}\right)-\Lambda\mathrm{_{coll}}n_\mathrm{H_2}e_\mathrm{c}\label{CR_diffusion_eq_basic},\\
    &\mathbb{K} \equiv \mathrm{K}_{i j}=\mathrm{K}_{\perp} \delta_{i j}+\left(\mathrm{K}_{\|}-\mathrm{K}_{\perp}\right) n_i n_\mathrm{j},\\ &n_i=\frac{{B_i}}{|\boldsymbol{B}|},
\end{align}
where $e_{\mathrm{c}},\,p_{\mathrm{c}},\, \mathbb{K},\Lambda_\mathrm{coll}$ and $n_\mathrm{H_2}$ are the CR energy density, CR pressure defined by $p_{\mathrm{c}} \equiv e_{\mathrm{c}}/3$, diffusion coefficient tensor, attenuation function due to collisions with gas, and number density of gas defined by $n_\mathrm{H_2}\equiv \rho/m_{\mathrm{H_2}}$, with $m_{\mathrm{H_2}}$ denoting the mass of a hydrogen molecule. Here we also adopt the ``single energy" approximation, i.e., all CRs are protons with a single CR kinetic energy $E_\mathrm{k}$. In addition, we ignore the advection and compression terms related to the gas velocity because these terms are smaller than the other terms as seen in Appendix \ref{Appendix_Advection_and_compression_term_in_diffusion_equation}. Thus, we treat the CR transport using the following anisotropic diffusion equation:
\begin{align}
    &\frac{\partial}{\partial t} e_{\mathrm{c}}=\nabla \cdot\left(\mathbb{K} \nabla e_{\mathrm{c}}\right)-\Lambda\mathrm{_{coll}}n_\mathrm{H_2}e_\mathrm{c}.\label{CR_diffusion_eq}
\end{align}
We solve this diffusion equation using the multigrid method (Appendix \ref{MGSolver}).

The total ionization rate $\zeta$ is given as
\begin{align}
    \zeta = \zeta_\mathrm{p}+\zeta_\mathrm{SLR}.\label{zeta_calculate}
\end{align}
Because the timescales of the CR diffusion and attenuation are both very short, the CR energy distribution quickly reaches a quasi-steady state within every hydrodynamic timestep (Section~\ref{Sec:the_difference_due_to_CR_energy}, see also Section 3.5 of \citet{Rodgers-Lee2017}). Therefore, we calculate the ionization rate by CR proton $\zeta_\mathrm{p}$ from the attenuation term in the CR diffusion equation as 
\begin{align}
    \zeta_\mathrm{p} = \frac{\Lambda_\mathrm{coll}e_\mathrm{c}}{\epsilon},\label{ec_zeta}
\end{align}
where $\epsilon = 50\,\mathrm{eV}$ is the average energy lost by each proton per ionization event \citep{Armillotta2021}. $\zeta_\mathrm{SLR}$ is the ionization rate by decay of short-lived radioactive nuclides (mainly $\mathrm{^{26}Al}$) and we adopt $\zeta_\mathrm{SLR} = 10^{-18}\,\mathrm{s^{-1}}$ \citep{Umebayashi2009} uniformly.

\subsection{Models}
We simulate disk formation processes in collapsing molecular cloud cores. We model the initial molecular cloud cores with an unstabilized Bonnor-Ebert (BE) sphere \citep{Ebert1955,Bonnor1956}. The initial density profile is 
\begin{align}
    \rho(r) = 1.5 \rho_{\mathrm{BE}}(r)
\end{align}
where $\rho_{\mathrm{BE}}$ is the density profile of the critical Bonnor-Ebert sphere. The density outside the critical BE radius is the same as that on the surface of the core. With uniform temperature of $T=10\,\mathrm{K}$, the mass and radius of this cloud are $M = 2\, M_\odot$, and $R \sim 13,440\,\mathrm{au}$. The
initial free-fall time at the center of the cloud is $t_\mathrm{ff}\sim 8.17\times 10^{4}\,\mathrm{years}$. We run the simulations until the maximum temperature reaches $T=2,000\,\mathrm{K}$, just before the onset of the second collapse.

We initialize the cloud with solid-body rotation and uniform magnetic fields both aligned to the $z$-axis, and add 1\% $m=2$ perturbation to the rotation, i.e., the initial velocities are 
\begin{align}
    v_x&=-\Omega y\left[1+0.01\cos(2\phi) \right],\\
    v_y&=\Omega x\left[1+0.01\cos(2\phi) \right],\\
    v_z&=0,
\end{align}
where $\phi\equiv \tan^{-1}(y/x)$. The angular rotation speed is $\Omega = 1.16\times 10^{-13}\,\mathrm{s^{-1}}$ or $\Omega t_\mathrm{ff} = 0.3$ \citep{Belloche2013,Pineda2019}. The magnetic field strength is $32\,\mu\mathrm{G}$, and the corresponding mass-to-flux ratio normalized by the critical value of stability for a uniform sphere \citep{Mouschovias1976} is $\mu \equiv \frac{M/\Phi_\mathrm{B}}{\left(M/\Phi_\mathrm{B}\right)_\mathrm{crit}} = 2$, which is comparable to magnetization of observed molecular cloud cores \citep{Crutcher2012}.

The boundary conditions are reflective for the non-ideal MHD part and isolated for the self-gravity part. The velocity outside the initial BE cloud is set to zero. In order to avoid numerical difficulties, we set a floor in the gas density so that the Alfven velocity $v_\mathrm{A} = |{\bf{B}}|/\sqrt{4\pi\rho}$ does not exceed $100\, c_\mathrm{s}$ where $c_\mathrm{s} = 0.19\,\mathrm{km\, s^{-1}}$ is the sound speed of cold gas at $10$ K. Since the mass increase due to the floor is only about 3\% of typical disk mass, it has a negligible influence on the dynamics.

The diffusion coefficient tensor $\mathbb{K}$ and the attenuation function $\Lambda_\mathrm{coll}$ depend on the CR kinetic energy $E_\mathrm{k}$. In this work, we choose $E_\mathrm{k} = 30\,\mathrm{MeV}$ as a representative CR kinetic energy for the single-bin approximation, considering the peak of the local Galactic CR spectrum observed by Voyager \citep{Ivlev2015,Cummings2016,Stone2019,Padovani2022} and the shape of the CR loss function \citep{Padovani2020}.The values of $\mathbb{K}$ and $\Lambda_\mathrm{coll}$ with $E_\mathrm{k} = 30\,\mathrm{MeV}$ are 
\begin{align}
    &\mathrm{K_\|}=3.9\times 10^{26}\,\mathrm{cm^2\,s^{-1}},\label{eq:coeff_uncorrected}\\
    &\mathrm{K_\perp}=3.9\times 10^{24}\,\mathrm{cm^2\,s^{-1}},\\
    &\mathrm{\Lambda_{coll}}=3.3\times 10^{-14}\,\mathrm{cm^3\,s^{-1}}\label{eq:att_func}.
\end{align}
These diffusion coefficients are calculated by scaling the values calculated for higher CR particle energies by \citet{SMP07a}. $\Lambda_\mathrm{coll}$ is calculated using the Stopping and Range of Ions in Matter package \citep{Ziegler2010}. For details, see Appendix~\ref{CR_diff_coefficient}.

These diffusion coefficients are estimated assuming that CR particles are scattered by turbulent magnetic fields. However, as we see later, the above diffusion coefficients are too high and need to be corrected in dense regions, because the mean free path of CR particles is limited by attenuation due to ionization rather than the scattering by magnetic fields in such a situation. We correct the diffusion coefficients as
\begin{equation}
\mathbb{K}_{\rm corrected}=\mathbb{K}\times\min\left(1, \frac{\Sigma_0^2 \Lambda_{\mathrm{coll}}}{m_{\mathrm{H}_2}^2 n_{\mathrm{H}_2}\mathrm{K_\|}}\right),\label{eq:cr_correction}
\end{equation}
where $\Sigma_0$ is the typical column density scale of the attenuation by ionization, and we adopt $\Sigma_0=96\,\mathrm{g\,cm^{-2}}$ \citep{Umebayashi1981}. This correction recovers the expected exponential decay of the CR ionization rate in the high density region. Derivation of Equation~(\ref{eq:cr_correction}) is given in Appendix~\ref{Sec:CRDiffusion_Correction}.

The boundary conditions for the CR transport part are fixed so that the CR ionization rate near the boundaries is fixed to a typical ionization rate in the ISM. In this work, we adopt the value of the external ionization rate, $\zeta_\mathrm{ex} = 10^{-16}\,\mathrm{s^{-1}}$. This is higher than the conventional value of $10^{-17}\,\mathrm{s^{-1}}$ \citep{Spitzer1978} but is motivated by recent observations of star-forming clouds \citep[e.g.,][]{Pineda2024}.

We calculate three models with different CR transport models. In Models DD (damped diffusion) and UD (uncorrected diffusion), we solve the CR transport equation (\ref{CR_diffusion_eq}) with and without the correction (\ref{eq:cr_correction}), respectively. Model CT is for comparison with previous works, and we use a constant ionization rate of $\zeta=10^{-16}\,{\mathrm{s^{-1}}}$ instead of solving the CR transport.

The base computational box spans $(6.67\times 10^{4}\,\mathrm{au})^3$ and is resolved with $128^3$ cells. We adopt Cartesian coordinates, and we use adaptive mesh refinement (AMR) to resolve small scale structures formed in the collapsing clouds. We adopt the Jeans criterion \citep{Truelove1998} and resolve the local effective Jeans length $\lambda_J = \left[\pi (c_\mathrm{s}+v_\mathrm{A})^2/G\rho\right]^{1/2}$ with at least 16 cells \citep{Commercon2008}. The maximum refinement level is 13 and the finest resolution is $0.064\,\mathrm{au}$ at the end of the calculations. In order to check convergence of the simulations, we have performed an additional simulation with twice the higher resolution for a short term from a checkpoint file, and found no qualitative difference.

\section{Results} \label{Result}
\subsection{Time Evolution}\label{Sec:time_evolution}
\begin{figure*}[!t]
\centerline{
\includegraphics[clip, width=1.0\textwidth]{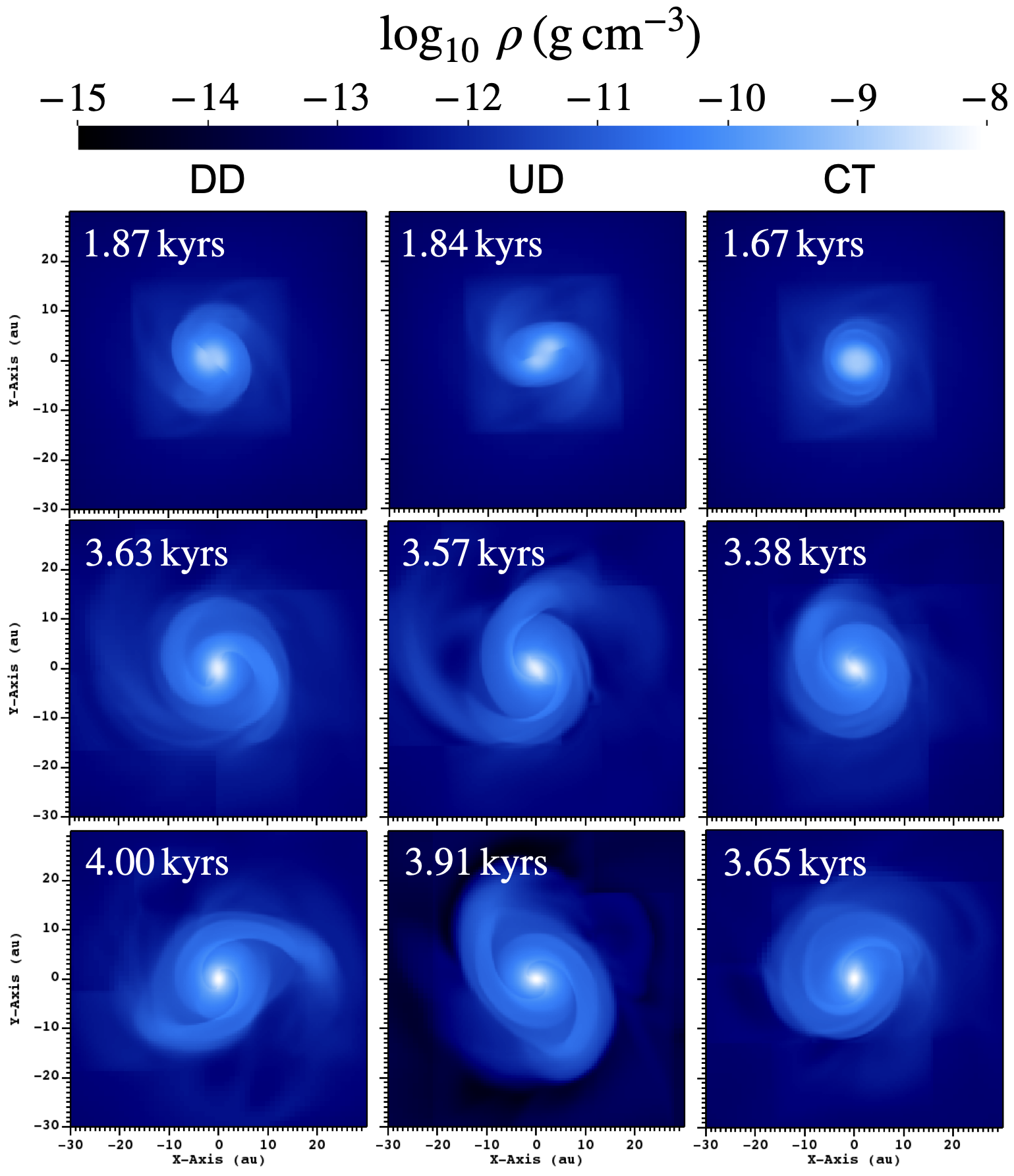}}
\caption{Horizontal cross sections of the gas density in the disk scale. The time elapsed since the first core formation is shown on the top left corner of each panel.
}
\label{fig:time_evolution_density}
\end{figure*}

\begin{figure}[htbp]
\centering
\includegraphics[width=0.48\textwidth]{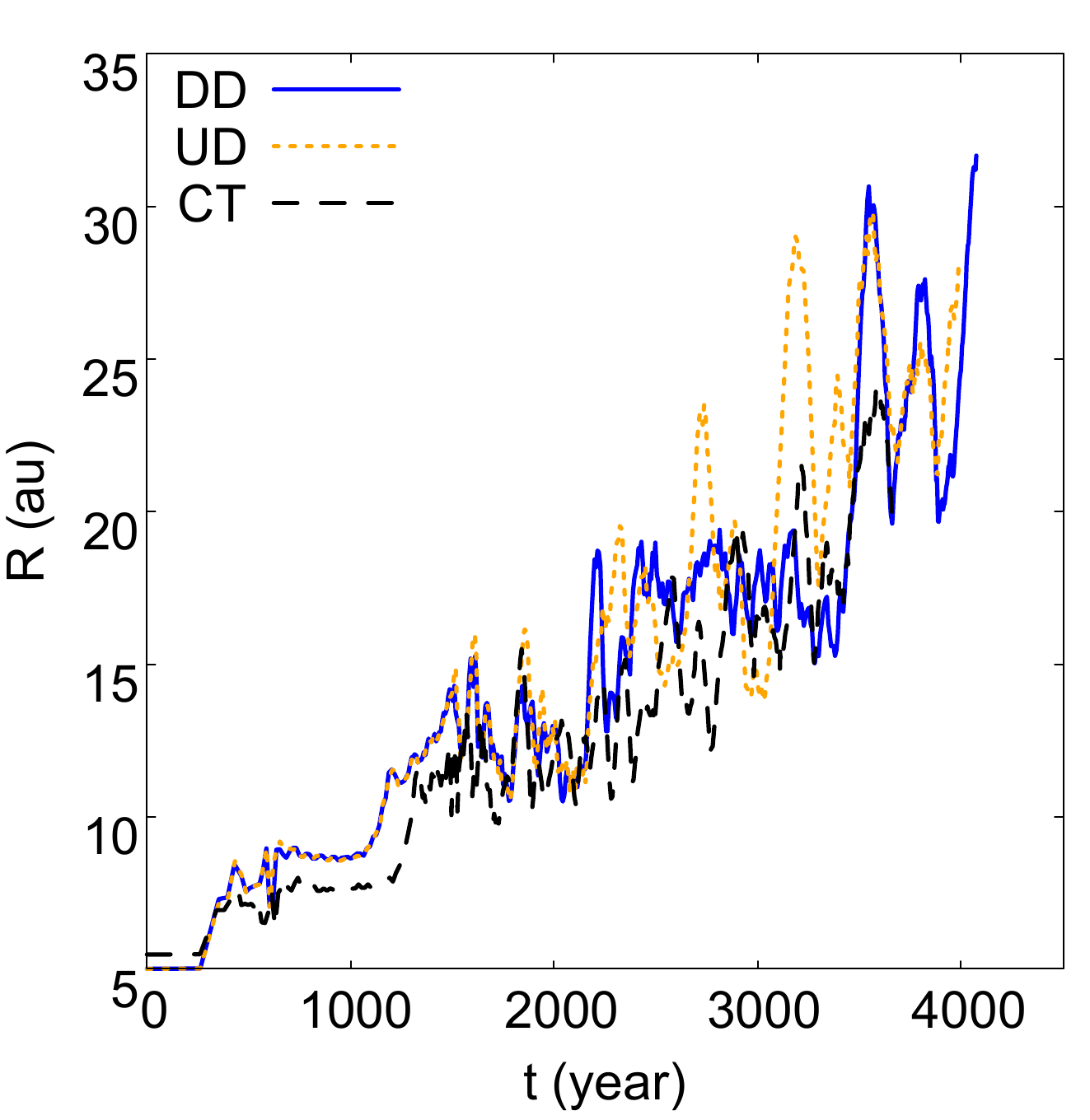}
\caption{Time evolution of the disk radii.}
\label{fig:t_radius}
\end{figure}

Figure \ref{fig:time_evolution_density} shows the evolution sequences of the gas density when the disk masses are $0.06\,M_\odot,0.1\,M_\odot$ and $0.11\,M_\odot$ from top to bottom, and the columns correspond to DD, UD and CT from left to right. Hereafter, we investigate the evolution of the disk after the formation of a first hydrostatic core, and we measure the time from the epoch when the maximum density exceeds $10^{-13}\,\mathrm{g\,cm^{-3}}$. Note that the square structures in the figure correspond to AMR level boundaries. Because the density plotted are measured at different heights with different resolutions, they are particularly visible where the vertical density gradient is steep.

Figure \ref{fig:time_evolution_density} indicates that the disks formed in DD and UD become larger and have more prominent spiral arms than in CT. This is because the angular momentum transport by the magnetic braking is more efficiently suppressed by the diffusion of magnetic fields and more angular momentum remains in the disks in DD and UD compared to the disk in CT. 

Figure \ref{fig:t_radius} shows the evolution of the disk radii. We define the disk as a region satisfying $\rho > 10^{-12}\,\mathrm{g\,cm^{-3}}$, $|v_\phi| > 2|v_R|$ and $v_z \mathrm{sign}(z)< 0.1\,c_s(10\,\mathrm{K})$, where $v_\phi,v_R$ and $v_z$ are the velocity components in cylindrical coordinates. The disks are initially small and gravitationally stable, but after $t > 1,000 \mathrm{\,years}$ the disks start growing because the disks become gravitationally unstable due to gas accretion from the envelope, and the spiral arms transport angular momentum within the disks by gravitational torque. The disk radii oscillate while growing because the spiral arms form and disappear recurrently \citep{Tomida2017}. The oscillation amplitude of the disk radii in DD and UD are greater than that in CT, indicating more intense gravitational instability. The difference in the disk radii between these models increases as the disks evolve. The disk radius in UD oscillates more significantly than in DD in the early phase, but the final radii are similar between the two models. 

\subsection{Disk Structure}\label{Sec:Comparision_disk}
\begin{figure*}[!t]
\centerline{
\includegraphics[clip, width=0.95\textwidth]{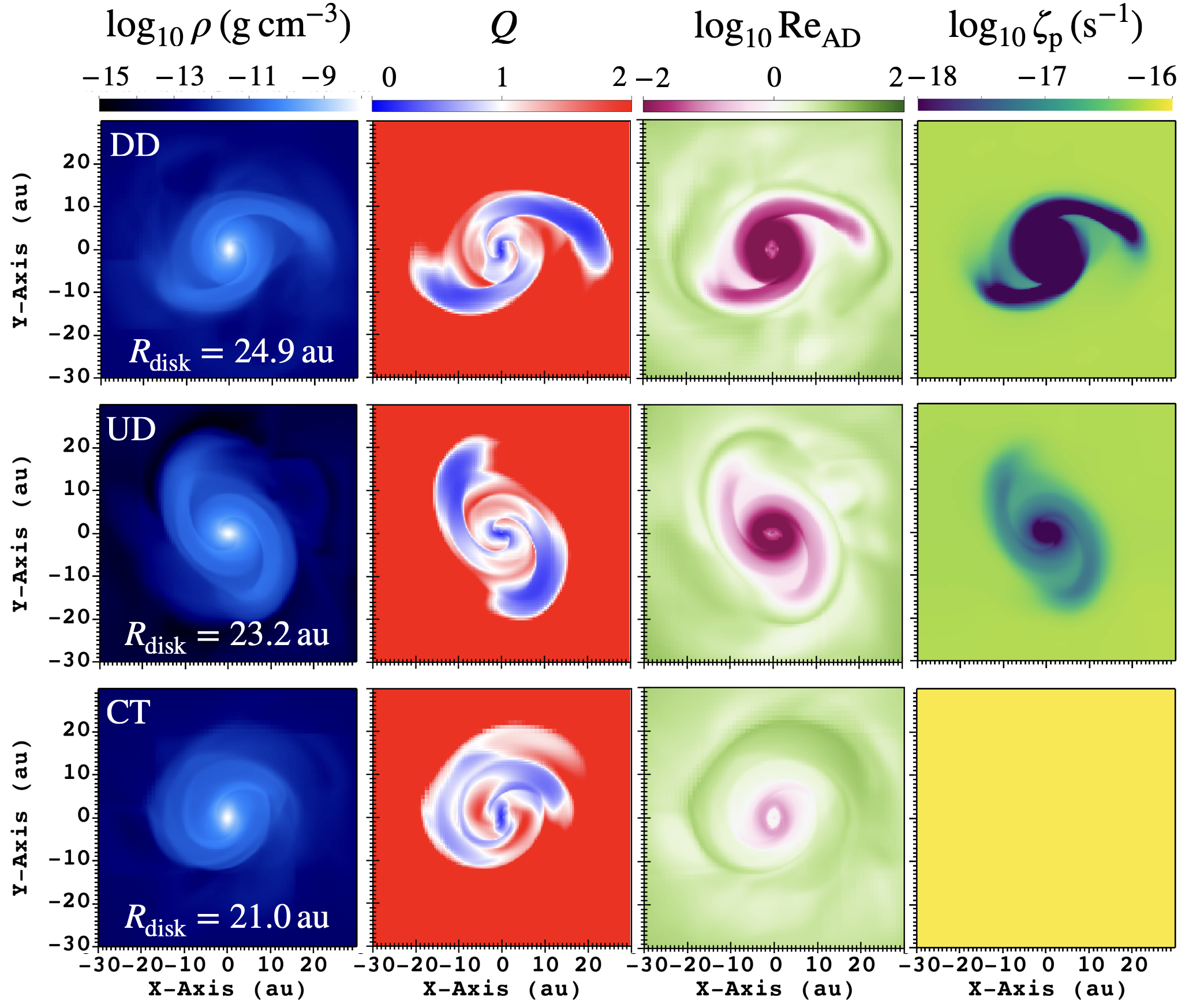}}
\caption{Distributions of the gas density, Toomre's Q value, Reynolds number of Ambipolar diffusion, and ionization rate by the CRs from left to right. While $\rho$, $\mathrm{Re_{AD}}$ and $\zeta_\mathrm{p}$ are cross sections on the disk midplane, $Q$ value is calculated by integration over the disk thickness as in Equation~(\ref{Qvalue}).}
\label{fig:disk_structure}
\end{figure*}

Figure \ref{fig:disk_structure} represents the gas density, Toomre's Q parameter \citep{Toomre1964}, Reynolds number of Ambipolar diffusion, and CR ionization rate when the disk mass is $0.11\,\mathrm{M_\odot}$. $Q$ is calculated as 
\begin{align}
     &Q \equiv \frac{\langle c_\mathrm{s}\rangle\langle\Omega\rangle}{\pi G\Sigma},\label{Qvalue}\\
     & \Sigma = \int ^h_{-h} \rho\, dz,\\
     & \langle c_\mathrm{s}\rangle = \frac{\int^h_{-h} \rho c_\mathrm{s}dz}{\Sigma},\\
     &\langle \Omega\rangle = \frac{\int^h_{-h} \rho \frac{v_\phi}{R}dz}{\Sigma},
\end{align}
where $\Sigma$ and $\Omega$ are the column density and angular velocity. The $\langle c_\mathrm{s}\rangle$ and $\langle \Omega\rangle$ are density-weighted average over a thickness of $2h$, and here we use $h=5\,\mathrm{au}$. While the disk thickness depends on radius, the results below are not sensitive to the choice of $h$ as we take density-weighted average which is strongly biased to the midplane values. $Q$ is smaller than unity in the spiral arms, indicating that they are indeed formed by the gravitational instability \citep{Takahashi2016}. The Reynolds number $\mathrm{Re_{AD}}$ is calculated as
\begin{align}
     &\mathrm{Re_{AD}} \equiv \frac{ \langle \boldsymbol{|v|}\rangle H }{ \langle\eta_\mathrm{AD}\rangle},\\
     &\langle \boldsymbol{|v|}\rangle = \frac{\int_h^{-h} \rho \boldsymbol{|v|}dz}{\Sigma}\\
     & H = \langle c_\mathrm{s}\rangle \langle \Omega \rangle^{-1}.
\end{align}
$\mathrm{Re_{AD}}$ is smaller than unity where magnetic diffusion is active. Since the Reynolds numbers are less than unity in most of the disks, the magnetorotational instability \citep[MRI, ][]{Balbus1991} is ineffective. Even if a certain region in the disks is not fully dead, MRI cannot grow because the characteristic wavelength of the fastest-growing MRI mode is greater than the disk thickness. Comparing the models between DD and CT, the disk in DD is clearly more gravitationally unstable as a result of more effective CR attenuation and stronger non-ideal MHD effects. The disks in DD and UD are similar because the Reynolds numbers are less than unity in the disk regions, and diffusion of magnetic fields is sufficiently effective in both models.

\begin{figure}[htbp]
\centering
\includegraphics[width=0.48\textwidth]{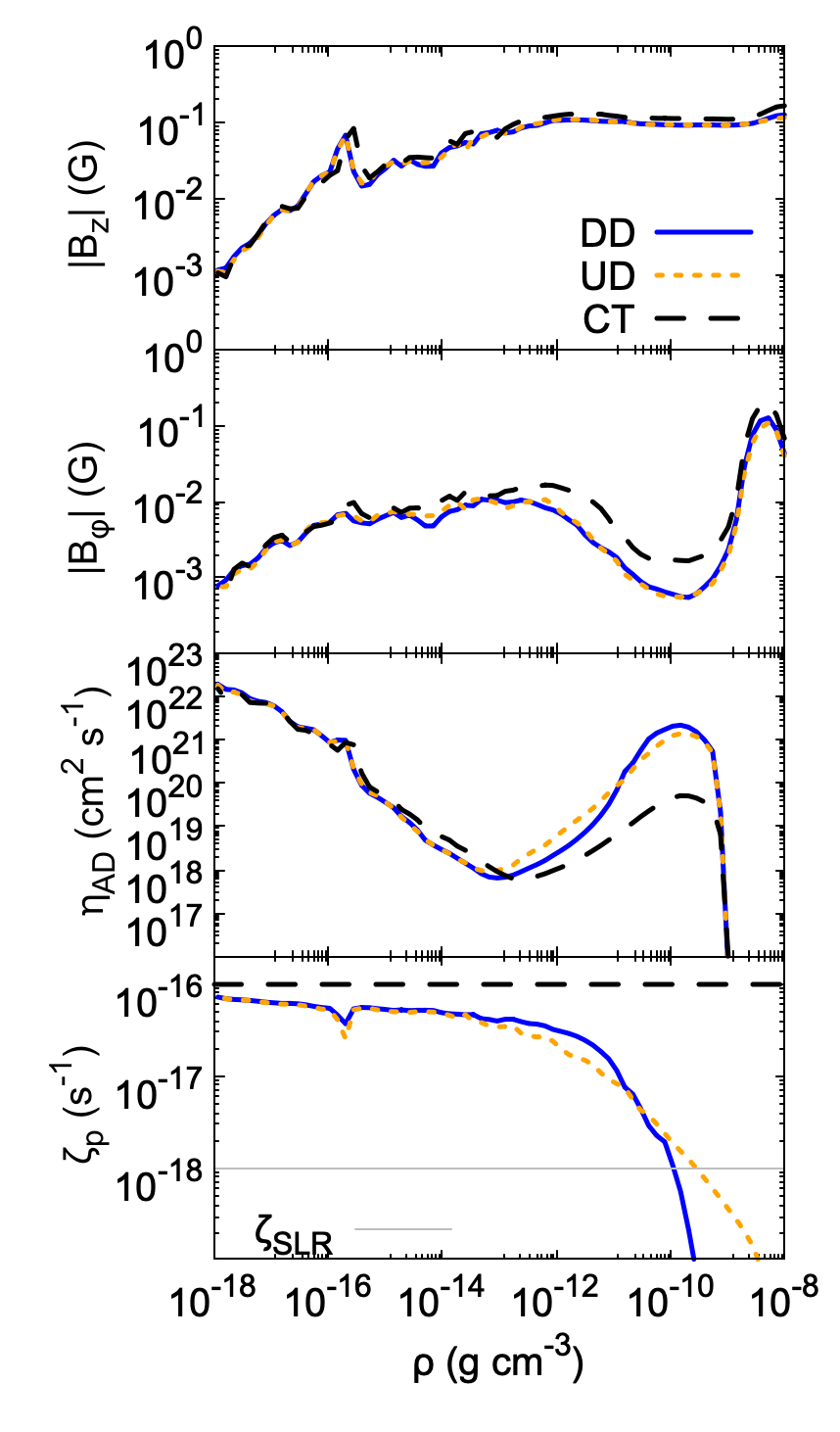}
\caption{Radial profiles of the magnetic fields, ambipolar diffusion rates and ionization rates by CRs. From top to bottom, $z$-component of magnetic flux density $|B_z|$, $\phi$-component of magnetic flux density $|B_\phi|$, ambipolar diffusion resistivity $\eta_\mathrm{AD}$, and ionization rate by CRs propagating from interstellar space $\zeta_\mathrm{p}$ when disk masses are $0.11\,\mathrm{M_\odot}$ . The gray line in $\zeta_\mathrm{p}$ panel shows the ionization rate by SLR $\zeta_\mathrm{SLR}$.}
\label{fig:rho_value}
\end{figure}

Figure \ref{fig:rho_value} represents radial profiles of z-component of magnetic flux density $|B_z|$, $\phi$-component of magnetic flux density $|B_\phi|$, magnetic resistivity of ambipolar diffusion $\eta_\mathrm{AD}$ and ionization rate by the CRs $\zeta_\mathrm{p}$ as functions of the gas density when the disk masses are $0.11\,\mathrm{M_\odot}$. We take density-weighted average of these values as 
\begin{align}
    A(R)\equiv \frac{\int_{-h}^h d z\int_0^{2\pi} d \phi  \rho(\mathrm{R}, \phi, \mathrm{z}) A(\mathrm{R}, \phi, \mathrm{z})}{\int_{-h}^h d z\int_0^{2\pi} d \phi \rho(\mathrm{R}, \phi, \mathrm{z})},
\end{align}
where $A(\mathrm{R}, \phi, z)$ is the arbitrary physical quantity to be averaged.

In the region outside the disk, where $\rho \lesssim 10^{-15}\,\mathrm{g\,cm^{-3}}$, $B_z$ behaves as $B_z \propto \rho^{2/3}$, indicating nearly spherical collapse with magnetic flux freezing. Inside the disk where $10^{-13}\,\mathrm{g\,cm^{-3}}\lesssim\rho \lesssim 10^{-9}\,\mathrm{g\,cm^{-3}}$, $B_z$ relaxes to a constant value because of efficient ambipolar diffusion \citep{Hennebelle2016}. In the central region of the disk with $\rho \gtrsim 10^{-9}\,\mathrm{g\,cm^{-3}}$, both $B_z$ and $B_\phi$ increase because the magnetic resistivities decrease sharply due to the thermal ionization of Potassium and the gas behaves almost like the ideal MHD again.

The ambipolar diffusion rates $\eta_\mathrm{AD}$ outside the disks are similar in all the models because the attenuation of the CR ionization rate is ineffective. However, $\eta_\mathrm{AD}$ in the disk regions are higher in DD and UD than that in CT because the CR attenuation is effective due to the high gas density. 
This difference affects the distribution of the magnetic fields. Inside the disks, the vertical magnetic flux $|B_z|$ is only slightly lower in DD and UD despite the higher diffusion rates, but the toroidal magnetic flux $|B_\phi|$ in DD and UD is smaller by a factor of a few than that in CT. We quantitatively compare angular momentum transport rates due to magnetic and gravitational torques in Section \ref{Sec:Angular_momentum_transport} and discuss the disk structure differences between the models in Section \ref{Sec:Disk structures and ionization rates by CRs}.

\subsection{Ionization Rate Distribution }\label{Sec:ionization_rate_distribution}

\begin{figure*}[!t]
\centerline{
\includegraphics[clip, width=0.8\textwidth]{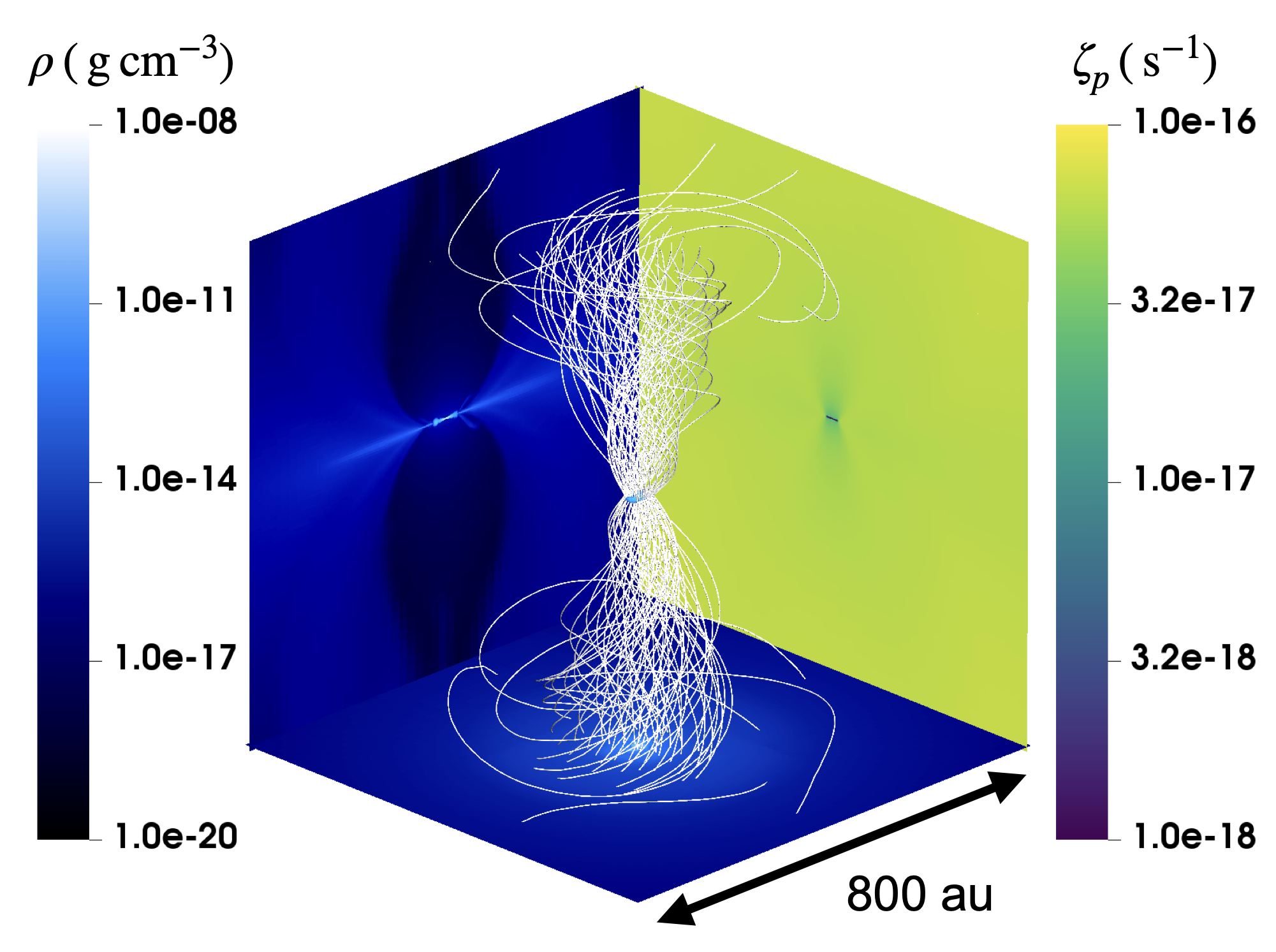}}
\caption{Density and CR ionization rate distribution for DD. The left and bottom walls of the box show the gas density profile $\rho \,(\mathrm{g\,cm^{-3}})$ on $y=0$ and $z=0$ planes, and the right wall shows the CR ionization rate profile $\zeta_\mathrm{p}$ on $x=0$ plane. White lines indicate magnetic field lines connected to the disk region.
}
\label{fig:3D_plot_rho_zeta}
\end{figure*}

Figure \ref{fig:3D_plot_rho_zeta} shows the density and CR ionization rate distributions in DD when the disk mass is $\mathrm{0.11\,M_\odot}$. The CR ionization rates decrease to about half of $\zeta_{\rm ex}$ in the envelopes outside the disks where $\rho < 10^{-13}\,\mathrm{g\,cm^{-3}}$ because the CR is attenuated by the gas in the envelope. In the central region of the disk, the CR ionization rate $\zeta_\mathrm{p}$ decreases by more than one order of magnitude and gets even below $\zeta_\mathrm{SLR}$. 
 
The magnetic field lines within the disk region are perpendicular to the disk surface because the non-ideal MHD effects are highly effective and straighten the field lines. The magnetic field lines above the disk are twisted by the rotation because the non-ideal MHD effects are ineffective and the magnetic fields are frozen in. The outflows magnetically driven from the disk open up the low-density cavities. The CR ionization rates in the outflow cavities are low despite the low gas density. This is because the cavities are connected to the disk with the magnetic field lines, and the CRs flow into the disk region and get absorbed there.

\begin{figure}[htbp]
\centering
\includegraphics[width=0.48\textwidth]{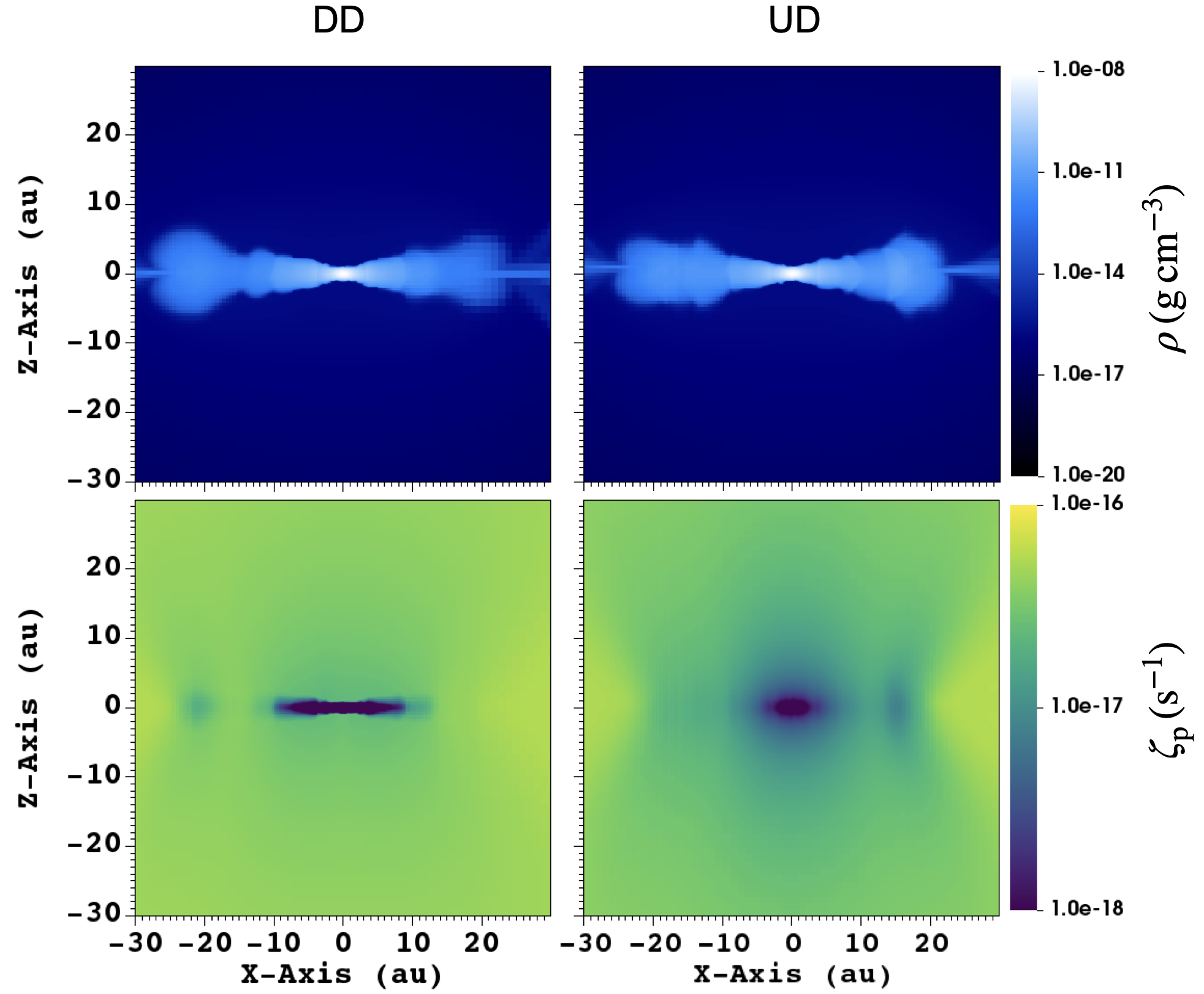}
\caption{Density and CR ionization rate distributions in DD and UD when the disk masses are $0.11\,\mathrm{M_\odot}$. For fair comparison, we take vertical cross sections containing the major axes of the disks.}
\label{fig:zeta_distribution}
\end{figure}

While the evolution and structures of the disks in DD and UD are similar, there is a visible difference in the disk size evolution around $t=3,000$ years in Figure \ref{fig:t_radius}. The larger oscillation amplitude in the disk radius indicates that the disk in UD is more unstable. In order to investigate this difference, we compare the CR ionization rate distributions between DD and UD in the disk scale in Figure \ref{fig:zeta_distribution}. The CR ionization rate in the midplane of the disk is lower in DD than in UD, but the region of low ionization rate in UD extends diffusely above the disk due to the higher diffusion coefficients. Therefore, the ionization rate on the disk surface in UD is lower than in DD, and the magnetic torque is suppressed slightly more strongly. As a result, the disk in the UD model becomes slightly more gravitationally unstable. However, this difference in evolution between DD and UD is minor, and both models eventually yield similar disk structures toward the end of the simulations.

\begin{figure}[htbp]
\centering
\includegraphics[width=0.48\textwidth]{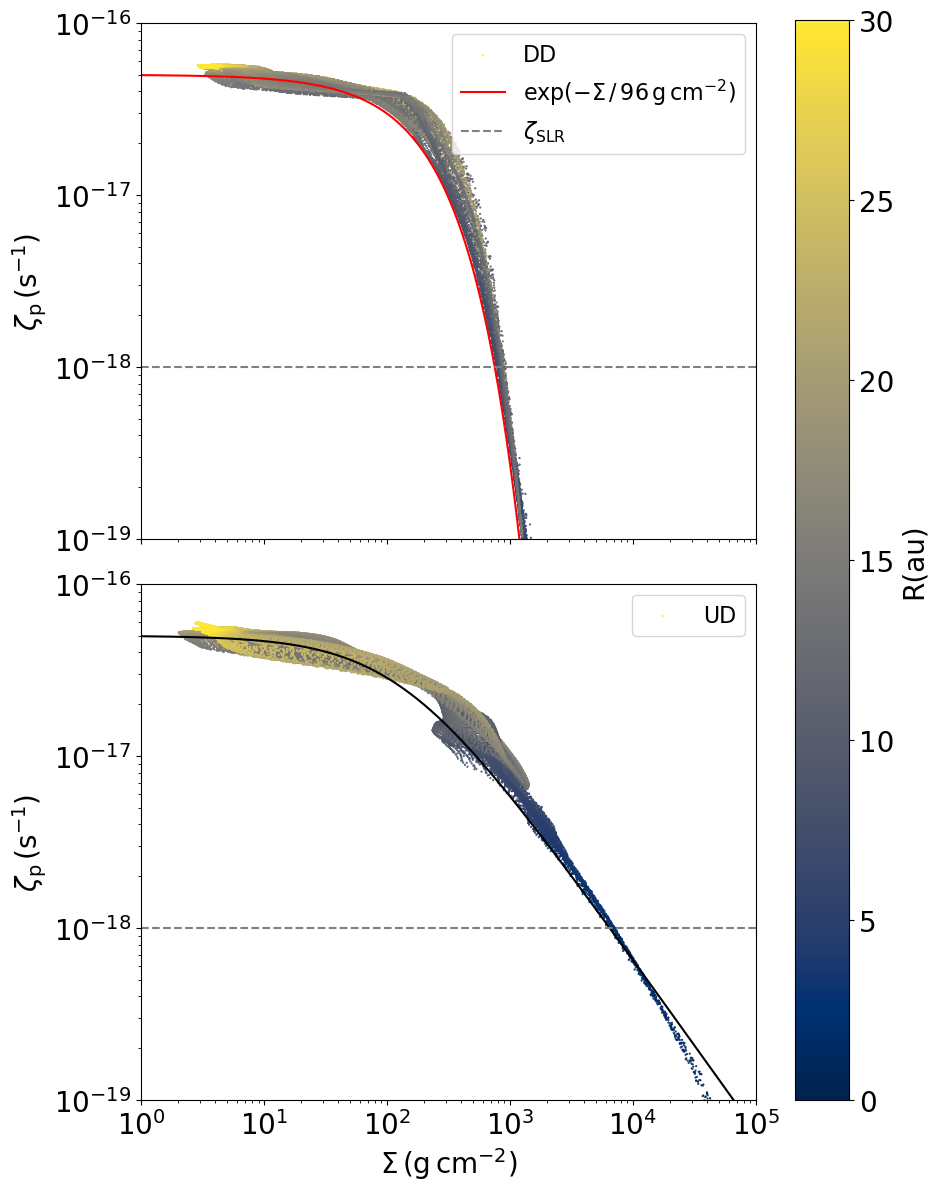}
\caption{Relation between the disk column density and CR ionization rate in the disk. From top to bottom, we show the model of DD and UD. The red line in the top panel is the model of \citet{Umebayashi2009}, while the black line in the bottom panel is a simple model based on the steady state of the uncorrected CR diffusion equation (Appendix~\ref{Appendix_Uncorrected_Diffusion}). The Gray dotted lines indicate the ionization rate by SLR $\zeta_\mathrm{SLR}$. The color of each point corresponds to the radius from the center of the disk.}
\label{fig:Sigma_zeta}
\end{figure}

Figure \ref{fig:Sigma_zeta} shows the relations between the disk column density and the ionization rate. We compute the disk column density by vertically integrating the density over the disk region as in Section~\ref{Sec:time_evolution} and show half of that as the horizontal axis. Because $B_\phi\ll B_z$ inside the disks, this is a reasonably good estimate of the column density along the magnetic field lines. On the vertical axis, we plot the minimum CR ionization rate in the integrated region. 
This figure demonstrates that the CR attenuation becomes significant beyond $\Sigma \gtrsim 10^2\, \mathrm{g\,cm^{-2}}$, which is consistent with previous studies on CRs in star-forming clouds \citep[e.g.][]{Umebayashi1981}. The CR ionization rate in DD exhibits exponential decay, but that in UD is a power-law decay. This behavior in UD is not physical, and the uncorrected diffusion equation overestimates the CR ionization rate in the high density gas. However, this behavior itself can be understood with a simple steady-state model which is shown with a black line in Figure \ref{fig:Sigma_zeta}. Derivation of this model is given in Appendix~\ref{Appendix_Uncorrected_Diffusion}.

\subsection{Angular Momentum Transport by Magnetic and Gravitational Torques}\label{Sec:Angular_momentum_transport}
\begin{figure}[htbp]
\centering
\includegraphics[width=0.48\textwidth]{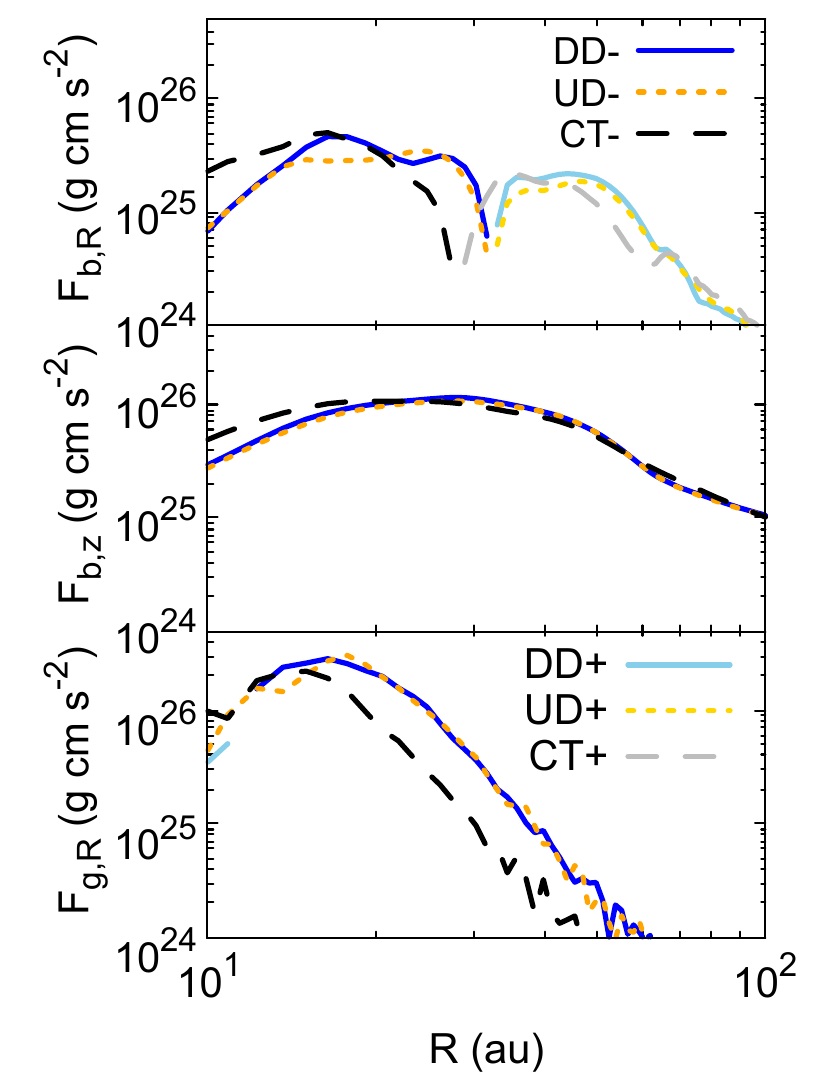}
\caption{Radial profiles of the angular momentum change rates due to radial transport by the magnetic torque $F_\mathrm{b,R}(R)$ (top), vertical transport by the magnetic torque $F_\mathrm{b,z}(R)$ (middle), and radial transport by the gravitational torque $F_\mathrm{g,R}(R)$ (bottom). Darker colored lines represent negative and lighter colored lines represent positive values.}
\label{fig:Angular_momentum_transport}
\end{figure}

To investigate the angular momentum transport in the different models, we quantify and compare the magnetic and gravitational torques. We compute radial distributions of angular momentum change rates by radial transport by magnetic fields, vertical transport by magnetic fields and radial transport by gravitational torque as:
\begin{align}
    F_\mathrm{b,R}(R,t) &\equiv \frac{1}{\Delta R}\int_{-h}^{h}\int_0^{2 \pi} \int ^{R+\Delta R/2}_{R-\Delta R/2} R dR d \phi dz \notag \\ 
    &\times   
    \frac{1}{R}\frac{\partial }{\partial R}  \left ( R^2 \frac{B_\phi(R, \phi, z) B_R(R, \phi, z)}{4\pi}\right),\\
    F_\mathrm{b,z}(R,t)&\equiv \frac{1}{\Delta R}\int_{-h}^{h}\int_0^{2 \pi} \int ^{R+\Delta R/2}_{R-\Delta R/2} R dR d \phi dz \notag \\ &\times
    \frac{\partial }{\partial z} \left ( R \frac{B_\phi(R, \phi, z) B_z(R, \phi, z)}{4\pi}\right) ,\\
    F_\mathrm{g,R}(R,t) &\equiv \frac{1}{\Delta R}\int_{-h}^{h}\int_0^{2 \pi} \int ^{R+\Delta R/2}_{R-\Delta R/2}  R dR d \phi dz\notag \\ &\times \frac{1}{R}\frac{\partial }{\partial R} \left ( -R^2 \frac{g_\phi(R, \phi, z) g_R(R, \phi, z)}{4\pi G}\right),
\end{align}
where $g_R$ and $ g_\phi$ are the gravitational acceleration $\boldsymbol{g}=-\boldsymbol{\nabla}\Phi$ in cylindrical coordinates, $h=5\mathrm{\,au}$ is the disk thickness and $\Delta R=1\,\mathrm{au}$ is the averaging interval. Figure \ref{fig:Angular_momentum_transport} shows $F_\mathrm{b,R}(R)$, $F_\mathrm{b,z}(R)$, and $F_\mathrm{g,R}(R)$ of each model. We take time average between the epochs shown in Figure~\ref{fig:disk_structure} and 1,000 years prior to them. 

The magnetic torque plays a dominant role outside the disks. In contrast, angular momentum transport by the gravitational torque dominates inside the disks. This indicates that these effects are complimentary; once the disks become dense and strongly decoupled from the magnetic fields by the non-ideal MHD effects, the disks become massive and gravitationally unstable, resulting in formation of spiral arms and angular momentum transport by the gravitational torque.

Comparing  DD and CT, the magnetic angular momentum transport is less efficient in DD because of the lower ionization in the disk region. In contrast, $F_\mathrm{g,R}(R)$ in DD is larger, indicating that the disk in DD is more gravitationally unstable than that in CT. This is the reason of the significant difference in the disks between the two models. The angular momentum transport flux in UD is slightly weaker than in DD, but the disks in these two models are similar. 

\section{Discussions}\label{Discussion}
\subsection{Ionization Rate Distributions and Their Influence on Disk Properties} \label{Sec:the_difference_due_to_CR_energy}
First, we estimate the gas density where the CR attenuation becomes significant by comparing the timescales of the CR diffusion $t_\mathrm{diff}$ and attenuation $t_\mathrm{att}$. We calculate these timescales as 
\begin{align}
    t_\mathrm{diff} &\sim \frac{H^2}{\mathrm{K_\|}} \notag\\
    &= 4.5\times 10^{-7} \left(\frac{H}{5\,\mathrm{au}} \right)^2   \left(\frac{\mathrm{K_\|}}{3.9\times 10^{26}\,\mathrm{cm^2\,s^{-1}}} \right)^{-1}   \,\mathrm{years},\label{t_diff}\\
    t_\mathrm{att}&\sim \frac{1}{\Lambda_\mathrm{coll}n_\mathrm{H_2}} \notag\\
    &= 3.2\times 10^{-5} \left(\frac{\Lambda_{\mathrm {coll}}}{3.4\times 10^{-14} \,\mathrm{~cm}^3\, \mathrm{~s}^{-1}}\right)^{-1} \notag\\ &\times
    \left(\frac{\rho}{10^{-13} \mathrm{\,g\,}\mathrm{~cm}^{-3}}\right)^{-1}\,\mathrm{years},\label{t_damp}
\end{align}
where the diffusion coefficient and attenuation function are shown in Equations~\ref{eq:coeff_uncorrected} and \ref{eq:att_func}.
The CRs attenuation is effective in the region where $t_\mathrm{diff} \gtrsim t_\mathrm{att}$, or $\rho \gtrsim \rho_\mathrm{damp}$, which is defined by
\begin{align}
    \rho_\mathrm{damp}&\equiv \frac{\mathrm{K_\|}}{\Lambda_\mathrm{coll}}\frac{m_\mathrm{H_2}}{H^2}\\
    & = 7 \times 10^{-12}  \left(\frac{\mathrm{K_\|}}{3.9\times 10^{26}\,\mathrm{cm^2\,s^{-1}}}\right) \notag\\ &\times\left(\frac{\Lambda_\mathrm{coll}}{3.4\times 10^{-14}\,\mathrm{cm^3\,s^{-1}}}\right)^{-1} \left(\frac{H}{5\,\mathrm{au}}\right)^{-2}\,\mathrm{g\,cm^{-3}}.\label{eq:rho_damp}
\end{align}
This value for the CR kinetic energy $E_k=30\mathrm{\,MeV}$ is comparable to the density in the outermost regions of the disks. Thus, the CR energy density is strongly attenuated within the disks, enhancing the non-ideal MHD effects (Figure~\ref{fig:rho_value}). In such regions, the angular momentum transport by magnetic fields is strongly suppressed, and the gravitational torque via the spiral arms becomes the dominant mechanism of the angular momentum transport (Figure~\ref{fig:Angular_momentum_transport}). 

We anticipate that the CR ionization rate distribution is sensitive to the choice of the representative CR kinetic energy. As the CRs with higher kinetic energy can penetrate deeper, we expect that the ionization degree remains higher and the disk size would be smaller if we use a higher value as the representative CR kinetic energy. In reality, CRs have an energy spectrum, and it is important to solve the distributions of CRs with different kinetic energies. It is expected that the ionization in low density regions are dominated by low-energy CR particles, and high density regions are predominantly ionized by high-energy CR particles. We also need to consider multiple CR species, not only protons but electrons and positrons produced as secondary particles \citep{Padovani2018}. Therefore, for realistic modeling, we need a multi-species multi-energy CR transport solver. This is an important direction of our future work.

In the very dense regions, the CRs are almost completely attenuated and the ionization rate is dominated by the decay of SLRs. In this work, we assume $\zeta_\mathrm{SLR}=10^{-18}\,\mathrm{s^{-1}}$ \citep{Umebayashi2009}. This value is obtained as a model for our solar system, where a substantial amount of SLRs ($^{26}$Al in particular) existed in the early phase of the formation. However, because the lifetimes of SLRs are comparable to the timescale of star formation ($\tau_{^{26}\mathrm{Al}}\sim 0.7$ Myr, for example), their abundances can vary between star-forming regions depending on nearby supernova activities. Therefore, our results infer that the disk size distributions may vary in various star-forming regions. On the other hand, if the CR attenuation is less effective (e.g., because of a harder CR spectrum), then environmental dependency of the disk size distributions can be less pronounced (yet it can depend on other environmental factors such as magnetic fields, turbulence, and metallicity) as Galactic CR distribution is ubiquitous (at least more ubiquitous than SLRs).

\subsection{Disk structure and Angular Momentum Transport}\label{Sec:Disk structures and ionization rates by CRs}
Here we discuss the origin of the difference in the formed disk structures such as the disk size and amplitude of the spiral arms between the models. As a result of CR propagation and attenuation, the distributions of the ionization rate and resulting ambipolar diffusion coefficient are different (Figure~\ref{fig:rho_value}). However, all the models show very similar $|B_z|$ distributions. This is because the $|B_z|$ is determined by balance between the magnetic flux accretion and outward diffusion by the non-ideal MHD effects, ambipolar diffusion in particular, at the outermost edge of the disks ($\rho\sim 10^{-13} -10^{-12}\,\mathrm{g\,cm^{-3}}$) \citep{Hennebelle2016}. In this region, the ambipolar diffusion coefficients are still similar because the CR attenuation reduced the ionization rates only moderately. Inside the disks ($\rho\gtrsim 10^{-12}\,\mathrm{g\,cm^{-3}}$), the ambipolar diffusion rates are different but they are sufficiently high, and the $|B_z|$ distributions anyway become constant as a consequence of efficient diffusion. Thus, all the models show similar vertical magnetic flux distribution.

The differences in the disk structures and evolution between the models originate from the difference in the toroidal magnetic fields inside the disks (Figure~\ref{fig:rho_value}). In the disks, the values of $B_\phi$ are determined by the balance between the ambipolar diffusion and the amplification by the disk rotation. Therefore, $|B_\phi|$ tends to be weaker in the models with lower ionization rates due to the CR attenuation. This effect suppresses the angular momentum transport by the magnetic fields in the disks (Figure~\ref{fig:Angular_momentum_transport}). The disks with the lower ionization rates become more massive and gravitationally unstable. This enhances angular momentum transport via gravitational torques, resulting in the larger disks (Figure \ref{fig:disk_structure}).

In this study, we neglect the Hall effect for simplicity. However, it is known that the Hall effect can significantly influence magnetic angular momentum transport \citep{Wardle2007,Tsukamoto2015_Hall}. Previous works have shown that when the magnetic field and the rotation vector are initially aligned, the magnetic angular momentum transport tends to be enhanced, resulting in a smaller disk. We will incorporate the Hall effect under realistic ionization rate distributions in future work.

\subsection{External Ionization Rate and Disk Formation}
\citet{Kuffmeier2020} showed that formation of protoplanetary disks are difficult in clouds with high ionization rates. Our results indicate that the CR ionization rate distribution is highly non-uniform due to the attenuation. As a result, even if the ionization rate in the surrounding ISM is high, the ionization rate in the disk scale can be significantly lower, and it can be as low as $\zeta_\mathrm{SLR}$. Therefore, our models predict that formation of protoplanetary disks are possible even in such regions with high ionization rates.

On the other hand, the ionization rate in the envelope scale is not significantly affected by the CR attenuation. Because the net magnetic flux brought into the disk scale is regulated around the outer edges of the disks, the dependence on the environmental CR ionization rate persists even in our models with the CR transport. We need more simulations covering long-term evolution and broader parameter space to predict the actual disk size distribution quantitatively, which are left for future works.

\subsection{Comparison with Recent Observations}
Recent Class-0 disk observations \citep[e.g.][]{Maury2019,Ohashi2023} have revealed many small disks of a few tens of au in size without prominent spiral arms \citep[but see also][]{Xu2023}. Our results indicate that even in star-forming regions with high ionization rates, large disks with prominent spiral arms can still form. While observational uncertainties should be carefully addressed, our results are apparently inconsistent with the observations. However, our simulations are only for the earliest phase of star formation, and the trend may change in the later evolution. As mentioned in Section \ref{intro}, once a central protostar forms, it can emit strong radiation which can warm up and ionize the disk gas. In addition, CRs can be accelerated at shocks in the accretion flow and/or jets \citep{Padovani2018,Margot2021} or by flares driven by magnetic reconnection in the disk \citep{Takasao2019,Kimura2023}. These processes may recover the ionization rate in the disk scale and enhance the magnetic angular momentum transport, resulting in disk shrinking.

Section \ref{Sec:ionization_rate_distribution} indicates that the CR ionization rate in the disk scale is highly non-uniform. This non-uniformity in the ionization rate may influence the chemical evolution in the disk scale. The ionization rate in the IM Lup protoplanetary disk derived from observations of $\mathrm{N_2H^+}$ and $\mathrm{H^{13}CO^+}$ is low in the inner disk and high in the outer disk, and the transition occurs near the spiral arm structure \citep{Seifert2021}. Our result is qualitatively consistent with this observation, as seen in Figure \ref{fig:disk_structure}. The low ionization rates are also reported in other protoplanetary disks such as HL Tau \citep{Pinte2016} and TW Hya \citep{Cleeves2015}. We should not compare our model and these observations naively because the disks in our model are much younger than these Class-II disks, but the observed low ionization rate in the inner disks can be explained by shielding of CRs.

On the other hand, some young stellar objects including B335 \citep{Cabedo2023} exhibit high ionization rates in the disk scale. The radius of the disk around B335 is notably small ($\lesssim 10\,\mathrm{au}$), and strong coherent magnetic fields are observed using dust polarization \citep{Maury2018}. These are consistent with efficient angular momentum transport by the magnetic fields. As the observed ionization rates around such objects is much higher than those in ISM, they may indicate existence of internal ionization sources. However, again, we need to carefully interpret these results as there are many sources of uncertainties.

\subsection{Validity of the CR diffusion approximation}\label{Sec:DiffusionApproximation}
In this work, we have adopted two major approximations for the sake of computational cost in order to enable time-dependent calculations of the ionization rate; the diffusion approximation and single energy approximation. As these approximations are not fully justified in the disk scale, here we discuss the limitations and potential impacts.

The diffusion approximation is a good approximation when the mean free path of CR particles is shorter than the scale of interest. From the diffusion coefficient we adopt, we can estimate the mean free path for pitch-angle scattering as
\begin{align}
  L_\mathrm{uncorrected}&\equiv\frac{3\mathrm{K_{\|}}}{v_\mathrm{cr}} \notag\\
  &= 10^4 \left(\frac{ \mathrm{K_{\|}} }{3.9\times 10^{26}\,\mathrm{cm^2\,s^{-1}} }\right)\left( \frac{v_\mathrm{cr}(E_\mathrm{k})} {7.6\times 10^9\mathrm{\,cm\,s^{-1}}}\right)^{-1}\,\mathrm{au}.
\end{align}
This is much larger than the disk scale and comparable to the cloud scale. Therefore, the (uncorrected) diffusion is a poor approximation. In the dense disk, however, the corrected diffusion coefficient including the effect of the attenuation leads to a mean free path comparable to the disk scale, given by 
\begin{align}
    L_\mathrm{corrected} &\equiv \frac{3\mathrm{K_{\|,corrected}}}{v_\mathrm{cr}} \notag\\
    &=24\left(\frac{\Sigma_0}{96\,\mathrm{g\,cm^{-2}} }\right)^2
    \left(\frac{\Lambda_\mathrm{coll}}{3.3\times 10^{-14}\,\mathrm{cm^3\,s^{-1}} }\right)\notag\\
    &\times \left(\frac{\rho}{10^{-10}\,\mathrm{g\,cm^{-3}}}\right)^{-1}
    \left( \frac{v_\mathrm{cr}(E_\mathrm{k})} {7.6\times 10^9\,\mathrm{\,cm\,s^{-1}}}\right)^{-1}\,\mathrm{au}.
\end{align}
Therefore, this approximation is at best marginally valid even with the correction, and we have to be aware of the limitation of the scheme and carefully interpret the results. We are planning to calculate CR energy distribution using a particle based method to trace CR trajectories and compare it with the diffusion approximation results.

As CRs flow along magnetic field lines, the CR energy density should increase where magnetic field lines converge. On the other hand, because the magnetic moment of a CR particle is an adiabatic invariant, CR particles with large pitch angles are reflected by converging magnetic fields. These effects, focusing and mirroring, cannot be fully taken into account in the diffusion approximation. While CRs flow along the field lines even in the anisotropic diffusion approximation, the CR energy density cannot be increased by the focusing effect because the energy is transported by its gradient. In our models, because the high density gas in the disks behaves as a sink of CRs, the CR energy flows from the ambient to the disks. Therefore, the focusing should be effective to some extent, but it is difficult to quantify. On the other hand, the mirroring effect is also not effective in the diffusion approximation, because it assumes that underlying CR distribution is always close to isotropic. This is appropriate if the pitch-angle scattering is efficient, but it is not likely to be the case as discussed above. In detailed calculations \citep{Silsbee2019,Fujii2022}, the focusing and mirroring effects are more or less balanced and cancel out. Therefore, our method may either over- or underestimate the CR ionization rate. However, as the CR distributions in the high density disks are dominated by the attenuation, our method with the corrected diffusion coefficients can reproduce reasonable ionization rate distributions in the disks and we expect that the disk evolution remains qualitatively similar even when more realistic CR transport model is incorporated. We are planning to compare our results with particle-tracing methods including these effects properly, but it is deferred to a future work.

It should be noted that the CR diffusion coefficients have large uncertainties. Because the low-energy CRs do not propagate a long distance and cannot be observed within the solar system due to the solar modulation, it is not possible to directly observe the diffusion coefficients (and production rate) of the low-energy CRs. Therefore, we simply adopted a theoretical model calculated for higher energies and scaled it to the lower energy (Appendix~\ref{CR_diff_coefficient}). However, there are two major sources of uncertainties. First, the diffusion coefficients for high energy CR particles already have a large uncertainty. We have adopted a value at $E_k = 10$ GeV from \citet{SMP07a} as a fiducial value, but there are other models based on different observations and assumptions \citep[e.g.,][]{2009ApJ...707L.179F,Genolini2019}. Second, we need to assume microscopic behavior of CRs in order to scale the diffusion coefficients to the lower energy, but it is not well constrained. We consider pitch-angle scattering by turbulent magnetic fields assuming the Kolmogorov spectrum, but the turbulence spectrum is highly uncertain in the small scale relevant to the low-energy CRs \citep[e.g.,][]{2009ApJ...707L.179F,Nava2013}. It can be weaker if the non-ideal MHD effects (e.g., ambipolar diffusion) work effectively and damp magnetic field perturbations, but it can be stronger if CRs themselves produce small-scale turbulence effectively by some instabilities \citep[e.g.,][]{Skilling1975,Bell2004,Shalaby2021}. If the diffusion coefficients are smaller, the CRs propagate slower and get attenuated more significantly, resulting in lower ionization degree in the disk scale. \citet{Hopkins2022b} point out that neither of the standard self-containment nor extrinsic turbulence model is compatible with observations. It is important to explore the impact of different diffusion coefficients on the disk structure and evolution, but it is beyond the scope of this paper.

\section{Conclusions}\label{Conclusion}
To study the effect of ionization by CRs on star and disk formation processes, we performed 3D nonideal MHD simulations coupled with CR transport. We focused on the early phase of star formation, and consider propagation and attenuation of the Galactic CRs. We summarize our findings as follows:
\begin{enumerate}
\item Considering the CRs transport, the ionization rate by CRs decreases significantly at the disk scale. Therefore, the disk is more gravitationally unstable and the disk size is bigger than that with constant ionization rate because magnetic angular momentum transport is suppressed.
\item Using the diffusion coefficient only considering scattering by magnetic turbulence leads to overestimation the mean free path and diffusion coefficient of CRs in dense regions, resulting in weaker attenuation of the CR ionization rate. With the correction incorporating the effect of attenuation (Equation~(\ref{eq:cr_correction})), we can reproduce the expected exponential decay of the CR ionization rate in dense regions.
\item The differences in evolution of disks arise from angular momentum transport due to variations in the ionization rates and resulting magnetic field distributions. The vertical magnetic fields brought into the disk do not differ substantially from model to model because they depend on the balance between the ambipolar diffusion and accretion rates near the outer edge of the disk. On the other hand, the toroidal magnetic fields differ and cause difference in angular momentum transport rates because the stronger the non-ideal MHD effects inside the disk are, the more amplification of the toroidal magnetic fields is suppressed.
\item Shielding of the Galactic CRs and low ionization rates in the disk scale can explain some observations of protoplanetary disks. However, we find many disks around young Class-0 protostars have small radii and few have pronounced spiral arms from surveys such as eDisk \citep{Ohashi2023} and CALYPSO \citep{Maury2019}. We may need to consider additional (internal) ionization source to explain those observations.
\end{enumerate}
 
In this work, we have adopted the single energy approximation for the CR transport, but in reality, CRs have an energy spectrum. As the penetration depth depends on the CR kinetic energy (higher-energy CRs can penetrate deeper), it is important to solve the spectrum at all positions. We are planning to extend our model so that multiple CR energies and species can be included.

Some observations and theoretical models suggest that CRs are accelerated in the vicinity of a protostar. We will include such an effect in our simulation, and extend our models to the late phase of star formation by performing long-term simulations.

\vskip\baselineskip
{\bf Acknowledgment} 
We thank Kazunari Iwasaki, Shinsuke Takasao and Shoji Mori for fruitful discussions. We are also grateful for the anonymous referee whose thoughtful comments greatly improved the manuscript. E.N. was supported by Graduate Program on Physics for the Universe in Tohoku University and Support from Japan Science and Technology Agency, Support for Pioneering Research Initiated by the Next Generation, Grant Number JPMJSP2114. This work was supported by Ministry of Education, Culture, Sports, Science and Technology (MEXT) and Japan Society for the Promotion of Science (JSPS) KAKENHI Grant Numbers JP22K0043 (K.T.), JP21H04487 (K.T. and S.S.K.), JP22K14028, JP23H04899 (S.S.K.), and JP24K17080(Y.K.). This research was also supported by MEXT as ``Program for Promoting Researches on the Supercomputer Fugaku" (Structure and Evolution of the Universe Unraveled by Fusion of Simulation and AI, JPMXP1020230406). K.T. acknowledges grant support from the Inamori Foundation. Numerical computations were carried out on Cray XC50 and HPE Cray XD2000 at the Center for Computational Astrophysics, National Astronomical Observatory of Japan, Genkai at the Research Institute for Information Technology, Kyushu University, Yukawa-21 at the Yukawa Institute Computer Facility, Kyoto University, and Camphor 3 at Institute for Information Management and Communication, Kyoto University. We used VisIt (Childs et al. 2012) to produce Figures \ref{fig:time_evolution_density}, \ref{fig:disk_structure}, \ref{fig:3D_plot_rho_zeta} \ref{fig:zeta_distribution}, \ref{fig:crtest}, \ref{fig:convergence}.

\appendix

\section{Advection and Compression Terms in the CR Diffusion Equation}\label{Appendix_Advection_and_compression_term_in_diffusion_equation}
We ignore the advection term $\nabla \cdot\left(e_{\mathrm{c}} \boldsymbol{v}\right)$ and compression term $p_{\mathrm{c}} \nabla \cdot \boldsymbol{v}$ related to gas motion in the diffusion equation (\ref{CR_diffusion_eq_basic}) because these terms are less significant than the diffusion term $\nabla \cdot\left(\mathbb{K} \nabla e_{\mathrm{c}}\right)$ and the attenuation term $\Lambda\mathrm{_{coll}}n_\mathrm{H_2}e_\mathrm{c}$. The timescales of the advection and compression terms are on the same order and can be estimated as 
\begin{align}
    t_\mathrm{adv} &\sim \frac{r}{v_K}\\
    &= \frac{r}{\sqrt{\mathrm{G}M / r}}=20\left(\frac{r}{5 \mathrm{\,au}}\right)^{1 / 2}\left(\frac{M}{0.01 \mathrm{\,M}_{\odot}}\right)^{-1 / 2} \,\mathrm { years },
\end{align}
and this timescale is far longer than $t_\mathrm{diff}$ (\ref{t_diff}) and $t_\mathrm{att}$ (\ref{t_damp}). Therefore, these terms do not have a significant effect on the ionization rate distribution in star-forming clouds.

\section{Multigrid CR Diffusion Solver} \label{MGSolver}
As the CR transport timescale is much shorter than dynamical timescales in star formation, we solve Equation~(\ref{CR_diffusion_eq}) fully implicitly. In this section, we provide brief description of the solver.
\subsection{Discretization}
Using the backward Euler discretization for the time derivative and centered differentiation for the spatial derivative, we obtain the following discretized equation.
\onecolumngrid
{\footnotesize
\begin{align}
\frac{e_{i,j,k}^{n+1}-e_{i,j,k}^{n}}{\Delta t}=\frac{1}{\Delta x}&\left[\frac{K_{xx}^{i+1/2,j,k}\left(e^{n+1}_{i+1,j,k}-e^{n+1}_{i,j,k}\right)-K_{xx}^{i-1/2,j,k}\left(e^{n+1}_{i,j,k}-e^{n+1}_{i-1,j,k}\right)}{\Delta x}\right.\notag\\
&+\left.{\frac{K_{xy}^{i+1/2,j,k}\left(e^{n+1}_{i+1,j+1,k}+e^{n+1}_{i,j+1,k}-e^{n+1}_{i+1,j-1,k}-e^{n+1}_{i,j-1,k}\right)-K_{xy}^{i-1/2,j,k}\left(e^{n+1}_{i,j+1,k}+e^{n+1}_{i-1,j+1,k}-e^{n+1}_{i,j-1,k}-e^{n+1}_{i-1,j-1,k}\right)}{4\Delta y}}\right.\notag\\
&+\left.{\frac{K_{xz}^{i+1/2,j,k}\left(e^{n+1}_{i+1,j,k+1}+e^{n+1}_{i,j,k+1}-e^{n+1}_{i+1,j,k-1}-e^{n+1}_{i,j,k-1}\right)-K_{xz}^{i-1/2,j,k}\left(e^{n+1}_{i,j,k+1}+e^{n+1}_{i-1,j,k+1}-e^{n+1}_{i,j,k-1}-e^{n+1}_{i-1,j,k-1}\right)}{4\Delta z}}\right]\notag\\
+\frac{1}{\Delta y}&\left[{\frac{K_{yx}^{i,j+1/2,k}\left(e^{n+1}_{i+1,j+1,k}+e^{n+1}_{i+1,j,k}-e^{n+1}_{i-1,j+1,k}-e^{n+1}_{i-1,j,k}\right)-K_{yx}^{i,j-1/2,k}\left(e^{n+1}_{i+1,j,k}+e^{n+1}_{i+1,j-1,k}-e^{n+1}_{i-1,j,k}-e^{n+1}_{i-1,j-1,k}\right)}{4\Delta x}}\right.\notag\\
&+\left.\frac{K_{yy}^{i,j+1/2,k}\left(e^{n+1}_{i,j+1,k}-e^{n+1}_{i,j,k}\right)-K_{yy}^{i,j-1/2,k}\left(e^{n+1}_{i,j,k}-e^{n+1}_{i,j-1,k}\right)}{\Delta y}\right.\notag\\
&+\left.{\frac{K_{yz}^{i,j+1/2,k}\left(e^{n+1}_{i,j+1,k+1}+e^{n+1}_{i,j,k+1}-e^{n+1}_{i,j+1,k-1}-e^{n+1}_{i,j,k-1}\right)-K_{yz}^{i,j-1/2,k}\left(e^{n+1}_{i,j,k+1}+e^{n+1}_{i,j-1,k+1}-e^{n+1}_{i,j,k-1}-e^{n+1}_{i,j-1,k-1}\right)}{4\Delta z}}\right]\notag\\
+\frac{1}{\Delta z}&\left[{\frac{K_{zx}^{i,j,k+1/2}\left(e^{n+1}_{i+1,j,k+1}+e^{n+1}_{i+1,j,k}-e^{n+1}_{i-1,j,k+1}-e^{n+1}_{i-1,j,k}\right)-K_{zx}^{i,j,k-1/2}\left(e^{n+1}_{i+1,j,k}+e^{n+1}_{i+1,j,k-1}-e^{n+1}_{i-1,j,k}-e^{n+1}_{i-1,j,k-1}\right)}{4\Delta x}}\right.\notag\\
&+\left.{\frac{K_{zy}^{i,j,k+1/2}\left(e^{n+1}_{i,j+1,k+1}+e^{n+1}_{i,j+1,k}-e^{n+1}_{i,j-1,k+1}-e^{n+1}_{i,j-1,k}\right)-K_{zy}^{i,j,k-1/2}\left(e^{n+1}_{i,j+1,k}+e^{n+1}_{i,j+1,k-1}-e^{n+1}_{i,j-1,k}-e^{n+1}_{i,j-1,k-1}\right)}{4\Delta y}}\right.\notag\\
&+\left.\frac{K_{zz}^{i,j,k+1/2}\left(e^{n+1}_{i,j,k+1}-e^{n+1}_{i,j,k}\right)-K_{zz}^{i,j,k-1/2}\left(e^{n+1}_{i,j,k}-e^{n+1}_{i,j,k-1}\right)}{\Delta z}\right]-\Lambda_{\rm coll}n_{\mathrm{H_2},i,j,k}e_{i,j,k}^{n+1}.\label{eq:cr_disc}
\end{align}}
\twocolumngrid
The diffusion coefficients at cell interfaces are evaluated as
\begin{align}
K_{xx}^{i+1/2,j,k}=\frac{K_{xx}^{i+1,j,k}+K_{xx}^{i,j,k}}{2},
\end{align}
etc. While the Constrained Transport solver of Athena++ solves face-centered magnetic fields, we use cell-centered magnetic fields to evaluate the diffusion coefficients for simplicity. Hereafter we assume $\Delta x = \Delta y = \Delta z = h$. We rewrite Equation~(\ref{eq:cr_disc}) into a matrix form as follows:
\begin{align}
\sum_{a=-1,0,1}\sum_{b=-1,0,1}\sum_{c=-1,0,1}M_{{i+a},{j+b},{k+c}}e^{n+1}_{i+a,j+b,k+c}=e^n_{i,j,k},\label{eq:matrix}
\end{align}
\onecolumngrid
where
\begin{align}
M_{i,j,k}=1+\Delta t&\left[\Lambda_{\rm coll}n_{\mathrm{H^2},i,j,k}+\frac{K_{xx}^{i+1/2,j,k}+K_{xx}^{i-1/2,j,k}+K_{yy}^{i,j+1/2,k}+K_{yy}^{i,j-1/2,k}+K_{zz}^{i,j,k+1/2}+K_{zz}^{i,j,k-1/2}}{h^2}\right],\notag\\
M_{i+1,j,k}=\Delta t&\left[-\frac{K_{xx}^{i+1/2,j,k}}{h^2}-\frac{(K_{yx}^{i,j+1/2,k}-K_{yx}^{i,j-1/2,k})+(K_{zx}^{i,j,k+1/2}-K_{zx}^{i,j,k-1/2})}{4h^2}\right],\notag\\
M_{i-1,j,k}=\Delta t&\left[-\frac{K_{xx}^{i-1/2,j,k}}{h^2}+\frac{(K_{yx}^{i,j+1/2,k}-K_{yx}^{i,j-1/2,k})+(K_{zx}^{i,j,k+1/2}-K_{zx}^{i,j,k-1/2})}{4h^2}\right],\notag\\
M_{i+1,j+1,k}=\frac{\Delta t}{4h^2}&\left(-K_{xy}^{i+1/2,j,k}-K_{yx}^{i,j+1/2,k}\right),\notag\\
M_{i-1,j+1,k}=\frac{\Delta t}{4h^2}&\left(+K_{xy}^{i-1/2,j,k}+K_{yx}^{i,j+1/2,k}\right),\notag\\
M_{i+1,j-1,k}=\frac{\Delta t}{4h^2}&\left(+K_{xy}^{i+1/2,j,k}+K_{yx}^{i,j-1/2,k}\right),\notag\\
M_{i-1,j-1,k}=\frac{\Delta t}{4h^2}&\left(-K_{xy}^{i-1/2,j,k}-K_{yx}^{i,j-1/2,k}\right),\notag\\
M_{i\pm1,j\pm1,k\pm1}=0,&\notag
\end{align}
\twocolumngrid
\noindent and the other matrix elements can be calculated analogously.

\subsection{Implementation}
Equation~(\ref{eq:matrix}) is a matrix equation with a 19-point stencil. We solve it using the multigrid solver based on \citet{Tomida2023}, which is originally developed for the Poisson equation of self-gravity. As the diffusion coefficients $\mathbb{K}$ and the absorption coefficient $\Lambda_{\rm coll}$ do not depend on the CR energy density, this equation is linear and the multigrid solver is applicable only with a few changes. For details of the base solver, we refer the reader to \citet{Tomida2023} and a textbook by \citet{Trottenberg}.

The first change is the smoother. While the Red-Black Gauss-Seidel smoother is used for self-gravity, it is not applicable to the present case because it has a larger stencil. Instead, we adopt the Red-Black Jacobi smoother, where we first apply the Jacobi smoother on red cells and then on black cells. In addition, we use the V(2,2) multigrid cycle, in which pre-smoothing and post-smoothing operations are applied twice. The other change is the interpolation at level boundaries. In self-gravity, we adopt a special interpolation formula at AMR level boundaries which conserves gravitational force lines. However, as the stencil contains transverse components, it is not straightforward to construct such a conservative formula for the CR diffusion. As a compromise, we simply apply the normal trilinear interpolation to calculate ghost cell at level boundaries. The other components such as the restriction and prolongation operators remain the same as the original solver. We apply the full multigrid (FMG) iteration in the first step, and then improve the solution by applying the V-cycle iterations until sufficient accuracy is achieved.

Note that the geometric multigrid solver using an isotropic smoother like the Jacobi smoother is not necessarily optimal for highly anisotropic diffusion problems. One can accelerate convergence using anisotropic semicoarsening and/or an anisotropic smoother like the plane Gauss-Seidel smoother if the direction of anisotropy is more or less global and known beforehand. However, it is difficult to apply such schemes for the present case, because the direction of anisotropy depends on magnetic fields, which is time- and position-dependent. It is also known that we can improve convergence by using an acceleration parameter $\omega$ larger than one. However, while we find that it indeed helps for purely diffusion problems with $\Lambda_{\rm coll}=0$, it becomes unstable in a very dense region when $\Lambda_{\rm coll}$ is not zero. This is because the CR energy density rapidly damps toward zero in such a dense region, and even small extrapolation can cause overshooting and result in a negative energy density. Therefore, we use $\omega=1$. Despite these compromises, our solver achieves sufficiently practical performance. We refer the interested reader to Chapter~5 of \citet{Trottenberg} for detailed discussions.

\subsection{Test Calculations} \label{sec:CRTest}
We demonstrate the capability of our solver using anisotropic diffusion of CRs in circular magnetic fields in 3D Cartesian coordinates similar to \citet{Parrish2005} and \citet{Jiang2018}. The computational domain is $[-1:1]^3$. The magnetic fields are fixed as
\begin{equation*}
B_x(x,y,z)= -\frac{y}{R}, \hspace{1em} B_y(x,y,z) = \frac{x}{R},    \hspace{1em} B_z(x,y,z) = 0,
\end{equation*}
where $R=\sqrt{x^2+y^2}$ is the cylindrical radius. The CR density is initialized as
\begin{equation*}
e_\mathrm{c}(x,y,z) = \begin{cases} 12 \; \left(\sqrt{(R-0.6)^2+z^2} < 0.1 \;\mathrm{and}\;|\phi|<\frac{\pi}{12}\right),\\
10 \;  (\mathrm{otherwise}),
\end{cases}
\end{equation*}
where $\phi$ is the azimuthal angle measured from the $x$-axis. In this test, the attenuation term is omitted. We set the diffusion coefficient along the magnetic field $K_\parallel=1$, and evolve the system to $t=0.26$ in one time step. Although this problem has an analytic solution when the diffusion is completely anisotropic (i.e. $K_\perp=0$), we do not compare our solutions with the analytic solution because our scheme cannot handle very strong anisotropy.

\begin{table}[t]
\begin{center}
\caption{CFL numbers and convergence rates for the anisotropic diffusion test.}
\label{table:convrate}
\begin{tabular}{l||r|r|r|r}
Resolution  & $1/32$ &  $1/64$ &  $1/128$ & $1/256$   \\
\hline
$\eta_{\mathrm{CFL}}$ & 133.1 & 532.5 & 2130 & 8520 \\
\hline
\hline
$K_\perp = 0.01$ & 0.818 & 0.852 & 0.884 & 0.902 \\
\hline
$K_\perp = 0.001$ & 0.885 & 0.906 & 0.942 & 0.964 \\
\hline
$K_\perp = 0.0001$ & 0.891 & 0.912 & 0.949 & 0.970 \\
\end{tabular}
\end{center}
\end{table} 

\begin{figure*}[t]
\centerline{
\includegraphics[width=0.78\textwidth]{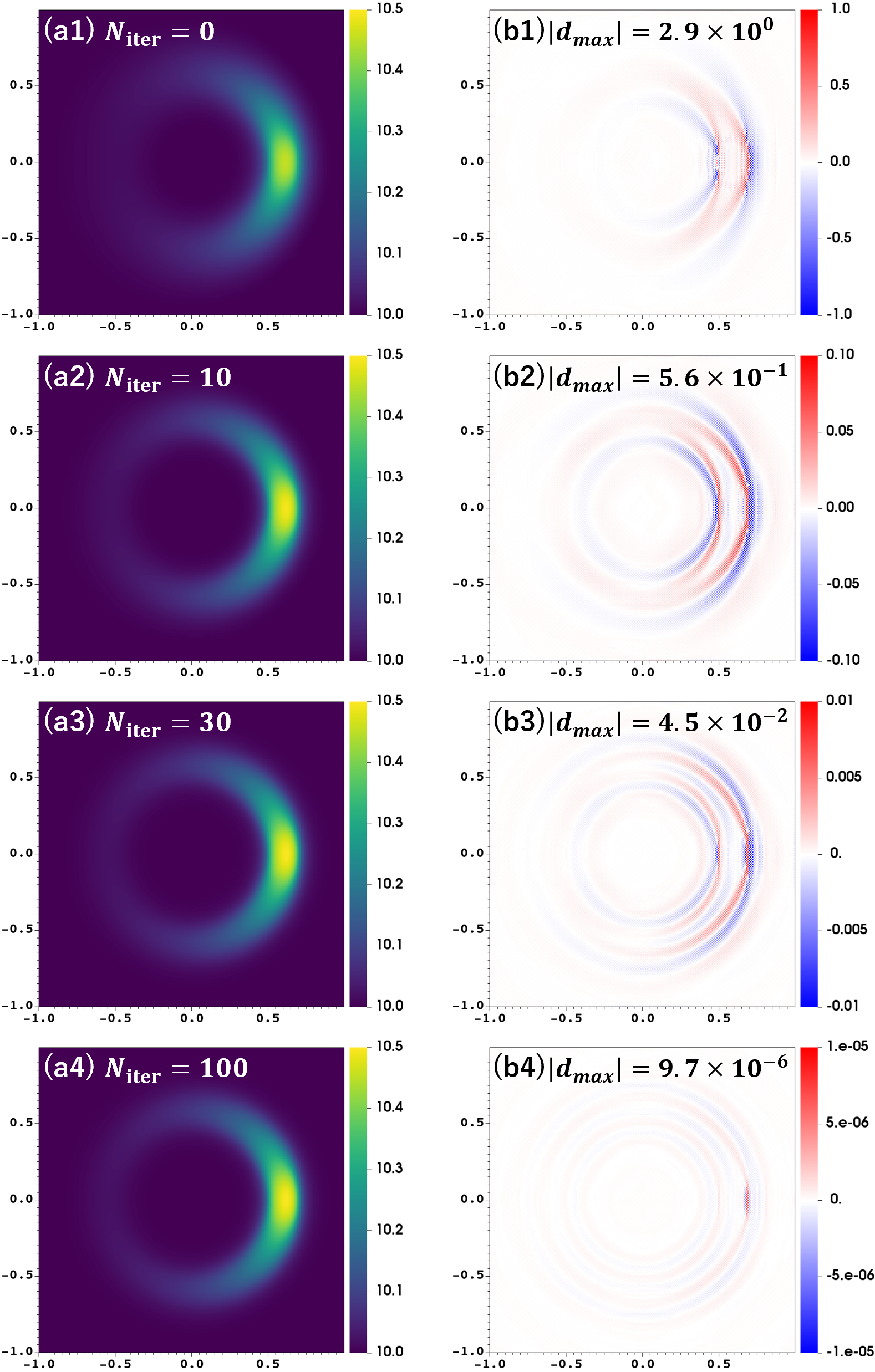}}
\caption{The solution (a1-a4) and defect (b1-b4) distributions in the $z=0$ plane of the anisotropic diffusion test with different numbers of iterations.
}
\label{fig:crtest}
\end{figure*}

\begin{figure*}[t]
\centerline{
\includegraphics[width=0.8\textwidth]{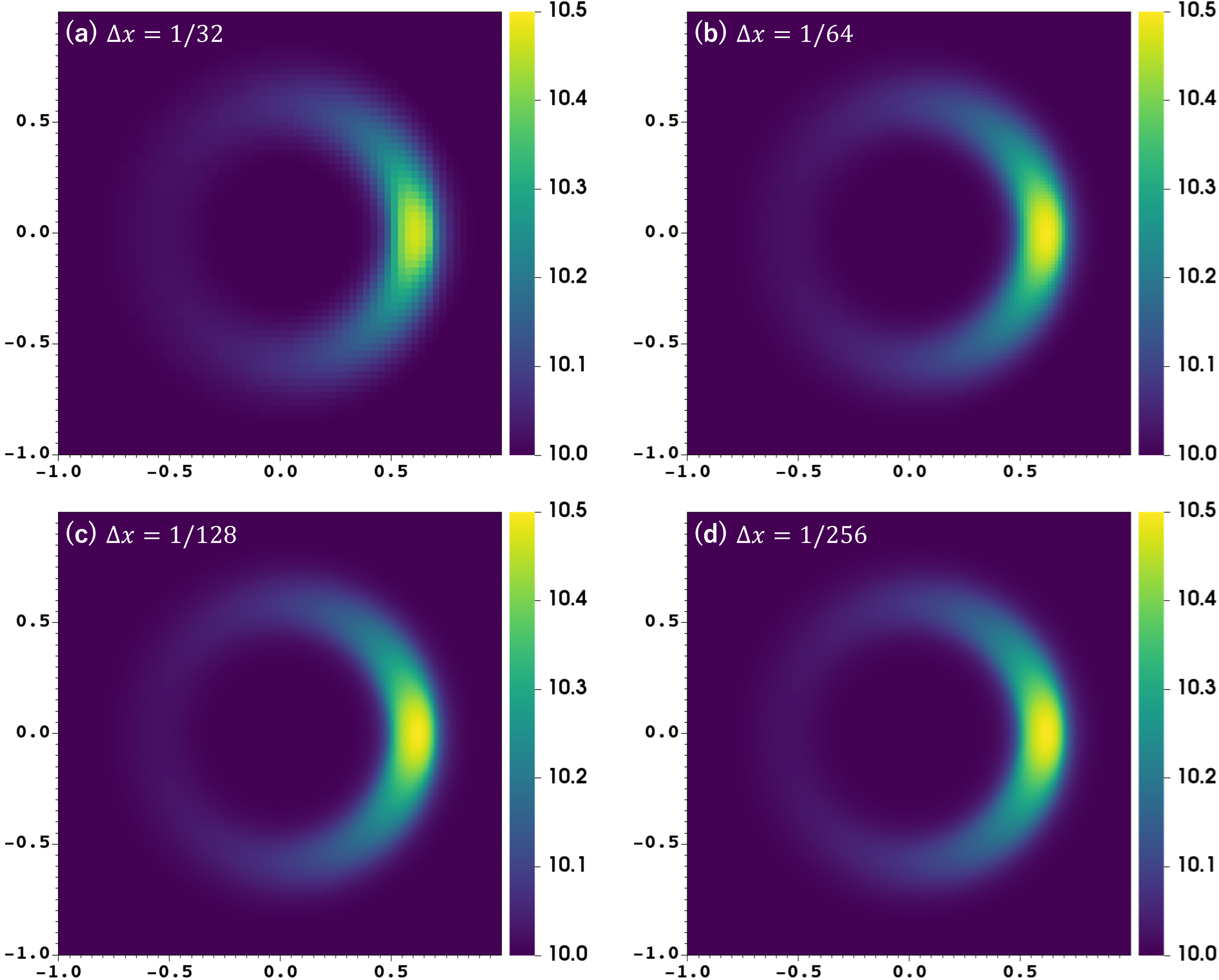}}
\caption{The anisotropic diffusion test results in the $z=0$ plane with different resolutions.}
\label{fig:convergence}
\end{figure*}

\begin{figure}[t]
\centerline{
\includegraphics[width=\columnwidth]{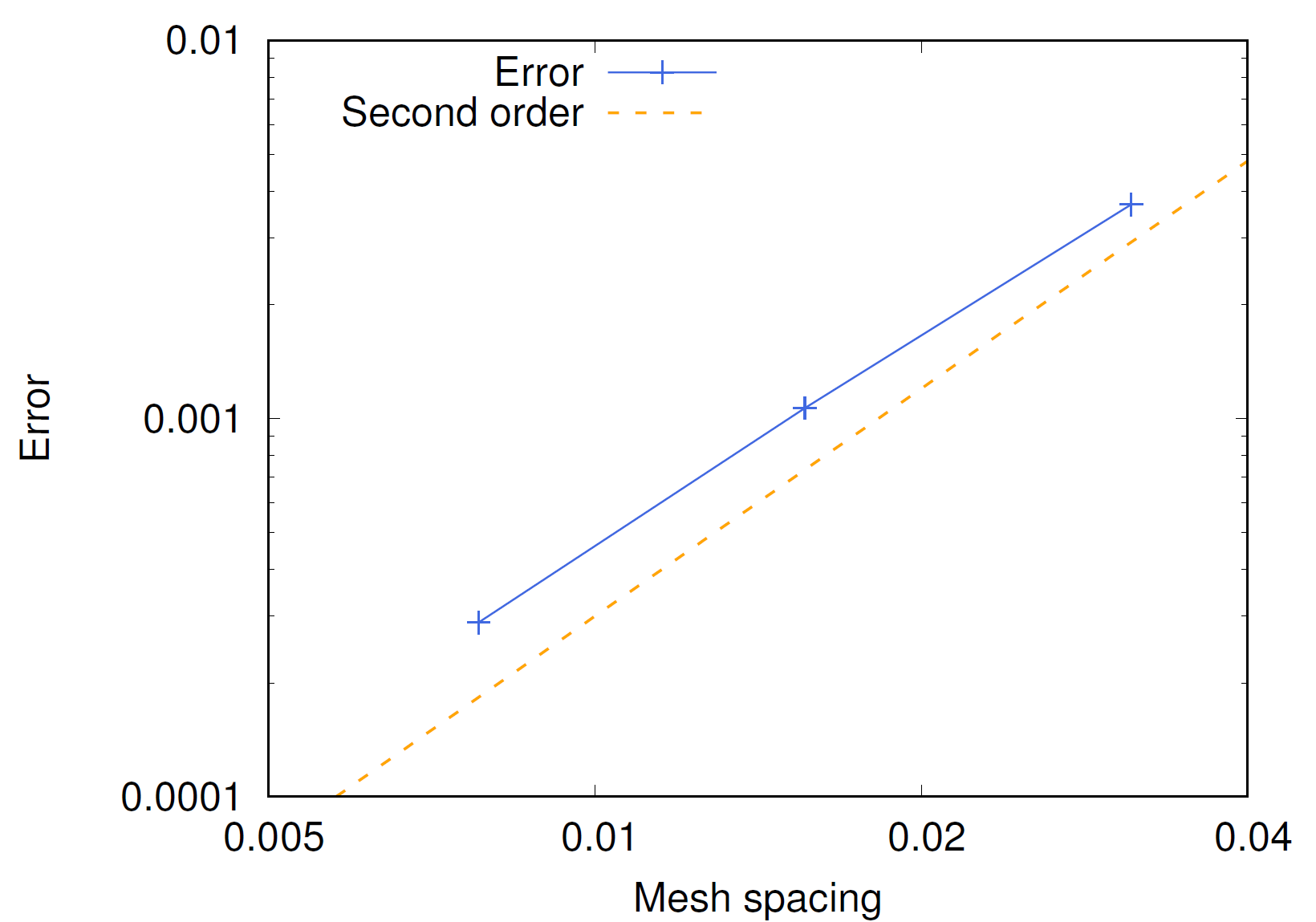}}
\caption{The error compared to the high resolution solution with $\Delta x=1/256$. The solver achieves almost second-order accuracy.}
\label{fig:convergence2}
\end{figure}

First, we measure the convergence rate, or the defect reduction factor per iteration, for different degrees of anisotropy. The average convergence rates in the first 100 iterations for various $K_\perp$ and different resolutions are shown in Table~\ref{table:convrate}. Note that the problem becomes more difficult with higher resolution, not only because of the larger degrees of freedom but also because of the larger Courant-Friedrichs-Lewy (CFL) number. The CFL number estimated for diffusion along the magnetic fields
\begin{equation}
\eta_{\mathrm{CFL}}\equiv\frac{K_\parallel \Delta t}{2\Delta x^2},
\end{equation}
is also shown in the Table. These convergence rates are not as good as those observed in the Poisson equation for self-gravity \citep{Tomida2023} because the multigrid solver with the isotropic smoother is not optimal for highly anisotropic problems. However, the solver still achieves practically useful performance as shown below.

Hereafter, we focus on the cases with $K_\perp = 0.01$. Figure~\ref{fig:crtest} shows the solution and defect distributions in the case with $\Delta x = 1/128$. Here we calculate the defect as 
{\small
\begin{align}
d_{i,j,k}^n\equiv&\left(\sum_{a=-1,0,1}\sum_{b=-1,0,1}\sum_{c=-1,0,1}M_{{i+a},{j+b},{k+c}}e^{n}_{i+a,j+b,k+c}\right)\notag\\-&e^n_{i,j,k}.\label{eq:defect}
\end{align}}
The initial FMG sweep ($N_\mathrm{iter}=0$) already gives a reasonably good initial guess. The chessboard patterns in the defects arise from the red-black Jacobi smoother. We find that 10-30 iterations are sufficient to achieve reasonable accuracies for practical applications. 

Figure~\ref{fig:convergence} compares the results with different resolutions using $N_{\mathrm{iter}}=100$. The solutions are consistent with the previous works \citep{Parrish2005, Jiang2018}.

Finally, we quantitatively measure the accuracy of the solver. In Figure~\ref{fig:convergence2}, we plot the volume-averaged error calculated as
\begin{equation}
\varepsilon(\Delta x)= \frac{\int |e_{c,\Delta x}-e_{c,\textrm{ref}}| dV}{\int dV}
\end{equation}
where $e_{c,\Delta x}$ is the solution obtained with resolution $\Delta x$ and $e_{c,\textrm{ref}}$ is the reference solution. Because the present test problem does not have an analytic solution, we use the high-resolution solution with $\Delta x = 1/256$ as the reference, and compare fully-converged solutions with lower resolutions against it. As shown in the Figure, the solver achieves almost second-order convergence, which is as expected from the central difference discretization of the diffusion operator (\ref{eq:cr_disc}).

\section{Diffusion Coefficients and Loss Function}\label{CR_diff_coefficient}
The CR diffusion coefficient is highly uncertain in star-forming region because many assumptions are required in the model and we cannot observe the CRs in star-forming regions. We extrapolate the values estimated at a higher energy in the Galactic ISM \citep[e.g.][]{SMP07a} to the low-energy assuming the Kolmogorov turbulence spectrum \citep{Kolmogorov1941}.  Although some study with gamma-ray data for a supernova remnant W28 suggests a lower diffusion coefficient in molecular clouds close to supernova remnants \citep[e.g.,][]{2009ApJ...707L.179F}, anisotropic diffusion model can explain the gamma-ray data of W28 with the typical Galactic cosmic-ray diffusion coefficient \citep{Nava2013}.

\begin{figure}[htbp]
\centering
\includegraphics[width=0.475\textwidth]{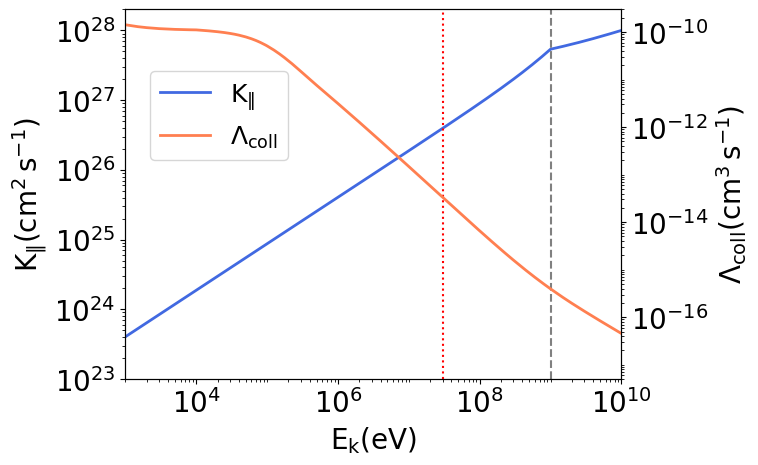}
\caption{The parallel diffusion coefficient and attenuation function of CR protons. The gray dashed line shows $E_\mathrm{k} = 1\mathrm{\,GeV}$ which is transition from non-relativistic to relativistic regimes. The red dotted line represents  $E_\mathrm{k} = 30\mathrm{\,MeV}$, which is adopted in this study.}
\label{fig:attenuation & diffusion function}
\end{figure}

We consider pitch-angle scattering caused by the resonance of turbulent magnetic fields and CRs in the direction along the magnetic field \citep{Jokipii1966,Schlickeiser2002,Kulsrud2005}. The diffusion coefficient in the direction along the background magnetic field by the pitch-angle scattering is estimated as 
\begin{align}
\begin{aligned}
& \mathrm{K}_{\|} \propto r_{\mathrm{g}} v_{\mathrm{cr}} F\left(k_{\mathrm{res}}\right)^{-1}, \\
& F\left(k_{\mathrm{res}}\right)^{-1} \equiv \int d k\left\langle\left(\frac{\delta B_k}{B_0}\right)\right\rangle^2\left|k_{\mathrm{res}}\right| \delta\left(k-k_{\mathrm{res}}\right),
\end{aligned}
\end{align}
where $r_{\mathrm{g}}, v_{\mathrm{cr}}$, and $F\left(k_{\mathrm{res}}\right)^{-1}$ are the gyro-radius, velocity of CR particles, and dimensionless power spectrum of the magnetic fields. The dimensionless power spectrum depends on the resonance wavelength $k_\mathrm{res}$ and the ratio of the perturbation magnetic field $\delta B_k$ to the background magnetic field $B_0$. Assuming that the magnetic turbulence has the Kolmogorov spectrum, the dimensionless power spectrum behaves as $F(k)^{-1}\propto k^{-2/3}$, so the diffusion coefficient scales as
\begin{align}
   \begin{split}
   & \mathrm{K}_{\|} (E_k)\\ = &
\begin{cases}
\mathrm{K}_{\|,0}(E_{k,0})\left(\frac{p_\mathrm{cr}\left(E_k\right)}{p_\mathrm{cr}(E_{k,0})}\right)^{1 / 3} \;(E_k \geq 1\mathrm{\,GeV}),\\
\mathrm{K}_{\|,0}(E_{k,0})\left(\frac{p_\mathrm{cr}(1 \mathrm{GeV})}{p_\mathrm{cr}(E_{k,0})}\right)^{1 / 3}\left(\frac{p_\mathrm{cr}\left(E_{\mathrm{k}}\right)}{p_\mathrm{cr}(1 \mathrm{GeV})}\right)^{4 / 3}\;(E_k \leq 1\mathrm{\,GeV}),
\end{cases}
\end{split}
\label{diffusion_coefficient_formula}
\end{align} 
where $\mathrm{K}_{\|,0}(E_{k,0})$ is the diffusion coefficient for a CR kinetic energy of $E_{k,0}$. We adopt $\mathrm{K}_{\|,0}(E_{k,0}=10\,\mathrm{GeV}) = 10^{28}\,\mathrm{cm^2\,s^{-1}}$\citep{Nava2013}, which is roughly consistent with the mean ISM value \citep[e.g.][]{SMP07a}. The dependence of the diffusion coefficient on CRs momentum changes from $\mathrm{K}_\| \propto p^{1/3} $ to $\mathrm{K}_\| \propto p^{4/3} $ at $E_k = 1\mathrm{\,GeV}$ switching from relativistic to non-relativistic. The result is shown in Figure \ref{fig:attenuation & diffusion function}.

The ratio between the parallel and perpendicular diffusion coefficients has large uncertainties, too. The quasilinear theory of diffusion coefficients (QLT: \citet{Jokipii1966,Schlickeiser2002}) predicts $\mathrm{K}_\perp \sim 10^{-6}\mathrm{K}_\|$, but test particle simulations with 3D turbulence indicate that the perpendicular diffusion coefficient is significantly higher, $\mathrm{K}_\perp \sim (0.02-0.04)\mathrm{K}_\|$ \citep{Giacalone1999}. We assume the diffusion coefficient perpendicular to the magnetic field $\mathrm{K}_\perp$ is 100 times smaller than $\mathrm{K}_\|$ in this work. 

The attenuation function $\Lambda_\mathrm{coll}$ is related to the CR loss function $L(E_k)$ as $\Lambda_\mathrm{coll}\equiv \frac{v_\mathrm{cr}(E_k)L(E_k)}{n_\mathrm{H_2}}$. We adopt the loss function from the Stopping and Range of Ions in Matter package \citep{Ziegler2010} \citep[see also][]{Padovani2020}, and the corresponding attenuation function is plotted in Figure~\ref{fig:attenuation & diffusion function}. 

\section{Correction of CR Diffusion Coefficients in High Density Region}\label{Sec:CRDiffusion_Correction}

If we naively use the diffusion coefficients above in the dense region, it causes overestimation of the diffusion speed and unphysically weak attenuation. This is because only scattering by turbulent magnetic fields are taken into account in the diffusion coefficient. In the dense region, attenuation by ionization is dominant and the mean free path of CR particles becomes significantly shorter. Here we derive a correction formula for the diffusion coefficients in order to reproduce a physically consistent limiting behavior in dense gas.

We assume a uniform slab of gas in a plane-parallel geometry, and the gas density is $n$. $z$ is the depth measured from the surface of the slab, and $\Sigma(z)=m_\mathrm{H_2}n_\mathrm{H_2}z$ is the corresponding column density. We also assume that magnetic fields are threading the slab and their directions are parallel to the slab normal, or $\boldsymbol{B}=B_0\boldsymbol{e}_z$. Ignoring diffusion perpendicular to the magnetic fields, it becomes a simple one-dimensional diffusion problem. Taking $\Sigma$ as the independent variable, the diffusion equation~\ref{CR_diffusion_eq} in a steady state reads:
\begin{equation}
\mathrm{K}_{\parallel}\frac{d^2e_c}{d\Sigma^2}=\frac{\Lambda_\mathrm{coll}}{m_\mathrm{n_{H_2}^2} n_\mathrm{H_2}}e_c.
\end{equation}
Assuming that the CR energy density diminishes in large $\Sigma$, the solution of this equation is 
\begin{equation}
e_c(\Sigma) = e_c(0)\exp\left(-\sqrt{\frac{\Lambda_\mathrm{coll}}{m_\mathrm{H_2}^2 n_\mathrm{H_2} \mathrm{K}_\parallel} }\Sigma \right).\label{eq:expdecay}
\end{equation}
This solution implies that the CR energy density decays exponentially in the uncorrected diffusion approximation, but its scale depends on the local gas density $n_\mathrm{H_2}$\footnote{This does not happen in the diffusion approximation of the radiation transport equation because the diffusion coefficient is inversely proportional to $\rho\kappa$ where $\kappa$ is the opacity, and the radiation energy decays exponentially as a function of the optical depth. This is because both diffusion and attenuation are caused by interaction with gas and dust in the case of radiation transport.}. This is why the uncorrected diffusion approximation exhibits slower decay in the high column density region (Figure~\ref{fig:Sigma_zeta}).  While this behavior is appropriate in low-column density regions, it is unphysical in high-column density (and high-density) regions. The CR energy density is expected to decay exponentially as a function of $\Sigma$ in the high density region where the attenuation is dominant. 

By analogy with radiation transport where the exponential decay scale does not depend on the gas density, we introduce correction for $\mathbb{K}$. Equation~(\ref{eq:expdecay}) can be rewritten as
\begin{equation}
    e_c(\Sigma) = e_c(0)\exp\left(-\frac{\Sigma}{\Sigma_0} \right),
\end{equation}
where
\begin{equation}
    \Sigma_0(E_k)\equiv\sqrt{\frac{m_\mathrm{H_2}^2 n_\mathrm{H_2} \mathrm{K}_\parallel(E_k)}{\Lambda_\mathrm{coll}(E_k)}},
\end{equation}
is the characteristic column density for the exponential decay. Without correction, $\Sigma_0$ can be artificially large in high density regions because the CR mean free path is solely set by scattering by turbulent magnetic fields. We limit $\Sigma_0$ in high density regions where the attenuation by ionization dominates the CR mean free path, so that we can reproduce the behavior obtained in more detailed calculations in the literature. Here we use
\begin{equation}
    \Sigma_{0,\mathrm{corrected}}(E_k) = \max (\Sigma_0, \Sigma_{0,\mathrm{max}}),
\end{equation}
with $\Sigma_{0,\mathrm{max}}=96\, \mathrm{g\, cm^{-2}}$  \citep{Umebayashi1981}. This is equivalent to Equation~(\ref{eq:cr_correction}). Note that this implementation is rather phenomenological, because the value of $\Sigma_0$ is calculated considering the energy spectrum and multiple species of CRs, but we apply it for a single CR particle energy.

\section{Asymptotic Behavior of the Uncorrected Diffusion Model} \label{Appendix_Uncorrected_Diffusion}
As Figure~\ref{fig:Sigma_zeta} shows, the uncorrected diffusion equation exhibits power-law decay of the ionization rate in the high density region. To understand this behavior, we conduct a simple estimate here. In a quasi-steady state, equation (\ref{CR_diffusion_eq}) reads 
\begin{align}
    \nabla \cdot\left(\mathbb{K} \nabla e_{\mathrm{c}}\right)-\Lambda\mathrm{_{coll}}n_\mathrm{H_2}e_\mathrm{c} =0.
\end{align}
To estimate the ionization rate near the midplane of the disk, we perform an order-of-magnitude estimate as
\begin{align}
    &\nabla \cdot\left(\mathbb{K} \nabla e_{\mathrm{c}}\right) \sim \frac{\mathrm{K_{\|} }\left(e_\mathrm{c,ex}-e_\mathrm{c,in}\right)}{l^2},\\
    &\Lambda\mathrm{_{coll}}n_\mathrm{{H_2}}e_\mathrm{c} \sim \Lambda\mathrm{_{coll}}n_\mathrm{H_2}e_\mathrm{c,in},\\
    &\Sigma \sim l m_\mathrm{H_2}n_\mathrm{H_2},
\end{align}
where $l,e_\mathrm{c,in}$ and $e_\mathrm{c,ex}$ are the length scale ($\sim$ disk thickness), the CR energy density near the disk midplane, and that outside the disk. Here we use the parallel CR diffusion coefficient as the magnetic field lines in the disk are almost vertical as a result of the non-ideal MHD effects. Since $e_\mathrm{c}\propto \zeta_p$, we obtain a simple relation between the ionization rate near the disk midplane and the column density of the disk is 
\begin{equation}
    \zeta_\mathrm{p,in} = \zeta_\mathrm{p,out}\left(\frac{2l\Lambda\mathrm{_{coll}} \Sigma}{m_\mathrm{H_2} \mathrm{K_{\|}}}+1\right)^{-1},\label{eq:stable_state}
\end{equation}
where $\zeta_\mathrm{p,out}$ is the ionization rate outside the disk. Note that $\zeta_\mathrm{p,out}$ is not the same as the ionization rate at the boundary of the computational domain $\zeta_\mathrm{p,ex}$ because of the attenuation in the envelope. Instead, we simply fit $\zeta_\mathrm{p,out}$ and $l$ to match the obtained distribution, and the result is shown as the black line in Figure~\ref{fig:Sigma_zeta}.


\bibliography{ms}{}
\bibliographystyle{aasjournal}



\end{document}